\documentclass[acmsmall,nonacm]{acmart}
\usepackage{hyperref}
\usepackage[most]{tcolorbox}
\usepackage{amsmath}
\usepackage{algorithm}
\usepackage{colortbl}
\usepackage{pgfplots}
\usepackage{algpseudocode}
\usepackage{graphicx}
\usepackage{textcomp}
\usepackage{xcolor}
\usepackage{booktabs}
\usepackage{array}
\usepackage{tikz}
\usetikzlibrary{positioning}
\usepackage{placeins}
\usetikzlibrary{arrows.meta}
\pgfplotsset{compat=1.17}
\def\BibTeX{{\rm B\kern-.05em{\sc i\kern-.025em b}\kern-.08em
    T\kern-.1667em\lower.7ex\hbox{E}\kern-.125emX}}
\def\TOOLNAME{RelicBOM}
\AtBeginDocument{%
  \providecommand\BibTeX{{%
    Bib\TeX}}}

\begin{document}

\title{\TOOLNAME{}{}: Recovering SBOM Components from ELF Binaries Under Incomplete Reference Coverage}

\author{Hanbo Yu}
\email{hanbo.yu@mail.mcgill.ca}
\authornotemark[1]
\affiliation{%
  \institution{McGill University}
  \city{Montreal}
  \country{Canada}
}

\author{Xinyu Hu}
\email{xinyu.hu@mail.mcgill.ca}
\affiliation{%
  \institution{McGill University}
  \city{Montreal}
  \country{Canada}}

\author{Zhiwei Fu}
\email{zhiwei.fu@mail.mcgill.ca}
\affiliation{%
  \institution{McGill University}
  \city{Montreal}
  \country{Canada}
}

\author{Steven H. H. Ding}
\email{steven.h.ding@mcgill.ca}
\correspondingauthor
\affiliation{%
 \institution{McGill University}
 \city{Montreal}
 \country{Canada}}

\author{Philippe Charland}
\authornotemark[2]
\email{philippe.charland@drdc-rddc.gc.ca}
\affiliation{%
  \institution{Mission Critical Cyber Security Section, Defence R\&D Canada}
  \city{Quebec}
  \country{Canada}}
\renewcommand{\shortauthors}{Yu et al.}

\begin{abstract}
Regulations increasingly require Software Bills of Materials (SBOMs), but SBOM generators read manifests and package databases that deployed binaries rarely retain. Tools that identify components in binaries instead match them against a reference, such as features compiled from known libraries or a catalogue of product names, and cannot report a component that their reference lacks. Vendor-only firmware components rarely appear in any reference, yet, to our knowledge, this limit has never been measured. We present \TOOLNAME{}, which recovers SBOM records from ELF binaries without requiring their components to appear in any reference. A language model proposes components and releases from evidence in the binary's strings, in one pass guided by a product catalogue and one open pass. A deterministic acceptance step reports a candidate only when the binary's own structure and strings support it, and states whether the component is embedded or a run-time dependency and which evidence supports it. On 1,173 Debian ELF files whose components lie outside the references of BLADE and CVE Binary Tool, \TOOLNAME{} recovers 39.6\% of the target components, whereas these two tools, EMBA, and Syft recover at most 0.3\%. It reports the correct release for 44.5\% of the components it recovers (89.3\% when one string carries both the name and the release), and 78.7\% of its records are correct. Most of this recall comes from the open pass; the acceptance step keeps the output to about one record per file, and a file costs less than half a US cent. On 101 proprietary firmware images, 148 of the 193 products that \TOOLNAME{} reports appear in the output of neither CVE Binary Tool nor EMBA.
\end{abstract}



\maketitle

\section{Introduction}
\label{sec:introduction}

In recent years, software supply chain security has emerged as a central concern across both industry and government. High-impact incidents, such as the SolarWinds compromise~\cite{solarwind} and the XZ backdoor~\cite{xz}, have shown that modern software is rarely developed in isolation. Instead, it is assembled from numerous open-source and proprietary components and redistributed across complex supply chains~\cite{ladisa2023sok}. This creates a fundamental software engineering problem: organizations must understand and track component-level composition even when software is deployed far from its original build environment. To support transparency, auditability, and downstream vulnerability analysis, the Software Bill of Materials (SBOM) has become a foundational artifact, driven by both operational risk management and strict regulatory mandates (e.g., US Executive Order 14028~\cite{14028} and EU guidelines~\cite{eu_SBOM}). Similar to bills of materials in manufacturing, SBOMs provide a machine-readable inventory of the components contained in a software artifact, together with attributes such as version, origin, and license.

SBOM generators read the metadata that builds and package managers leave behind, such as manifests, lockfiles, package databases, and container layers~\cite{SBOM_lr}. Even with that metadata, generators disagree on what an artifact contains and omit dependencies~\cite{yu2024sbomcorrectness,xiao2025jbomaudit}, and the choice of generator changes which vulnerabilities are reported~\cite{impact}. After deployment, the metadata itself is often gone. Stripped executables, repackaged firmware, and vendor bundles carry no manifest and no package database~\cite{empirical}. An auditor who receives the firmware of an infotainment unit or a network appliance typically has the binaries and nothing else. For these artifacts, the inventory must be recovered from the binaries themselves.

\begin{figure*}[t]
  \centering
  \includegraphics[width=\linewidth]{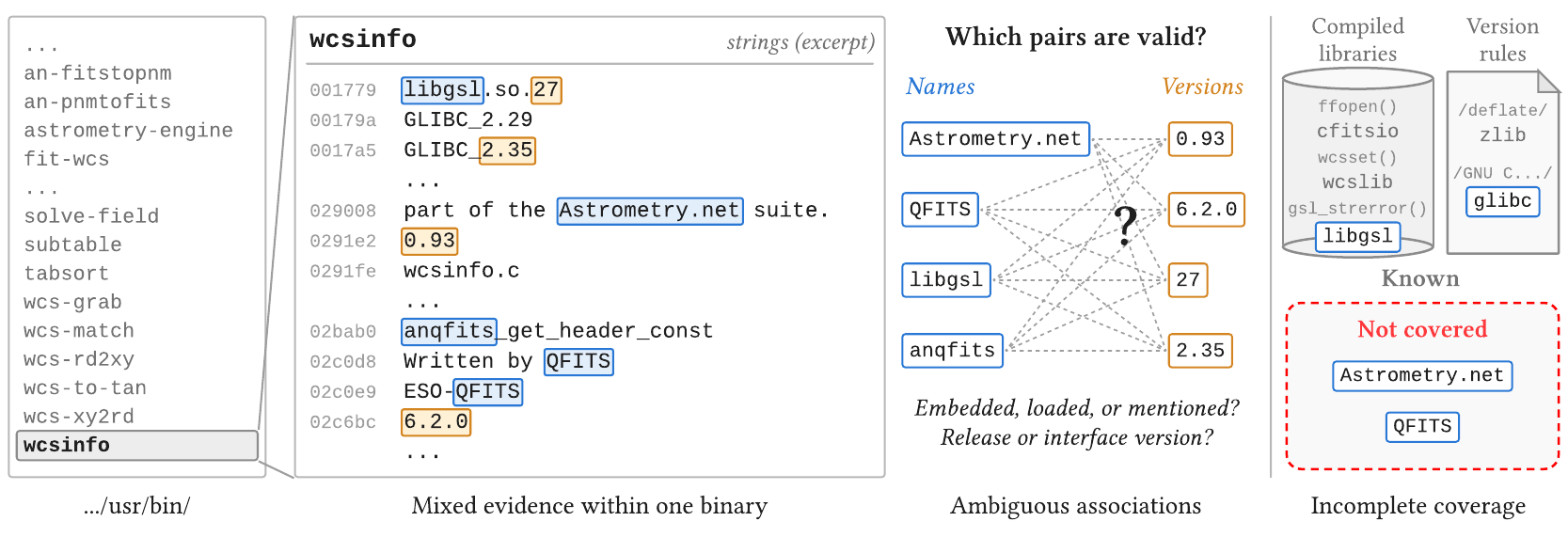}
    \caption{Component recovery on \texttt{wcsinfo} (Astrometry.net 0.93). Its strings mix component names (blue) with version-like tokens (orange) that only partly belong together. The knowledge of existing tools, features of compiled libraries and version rules, covers only the libraries the file loads, not its product or the library it embeds.}
  \label{fig:demo}
\end{figure*}

Binary-only component identification is an established line of work, and it rests on a common assumption: the tool must already know the component. Similarity-based detectors such as LibDB~\cite{tang2022libdb}, FirmSec~\cite{zhao2024firmsec}, BinaryAI~\cite{jiang2024binaryai}, and BLADE~\cite{liu2026blade} compile or index known libraries and match features of the result, such as function names, strings, and symbols, against the binary. Scanner-oriented tools such as CVE Binary Tool~\cite{cvebintool}, EMBA~\cite{emba}, and Syft~\cite{syft} instead pair a list of product names with version rules, hand-written patterns that match the release banner of each product. We call such knowledge a reference. When the target is in the reference, these tools can be very precise; BLADE, for example, reports high precision and recall on a benchmark whose libraries are all in its corpus~\cite{liu2026blade}. When it is not, they cannot report it. LibDB can only report reuse of libraries in its database and falls back to the closest version it knows~\cite{tang2022libdb}, and FirmSec states that components missing from its database become false negatives~\cite{zhao2024firmsec}. BLADE built its corpus to cover every library in its benchmark, and still reports a related library when the true one is not among its retrieved candidates~\cite{liu2026blade}. Because published evaluations construct their ground truth from libraries inside the reference, this blind spot has, to our knowledge, never been measured. Figure~\ref{fig:demo} shows it on \texttt{wcsinfo}, an executable of the Debian project Astrometry.net: both kinds of reference know the libraries that the file loads, GSL and glibc, but neither the product that the file belongs to nor the library QFITS that it embeds, although these are what its SBOM must state. The blind spot matters most where SBOMs matter most. Vendor-private components in firmware were never compiled into a public corpus and rarely have a version rule. In one infotainment image we analyzed, a shared library named \texttt{libIMGegl.so} suggests a graphics component, yet its strings include \texttt{qnx6.5.0-sp1}, which identifies the QNX Neutrino real-time operating system and its release. QNX is absent from the feature database we examined, and neither CVE Binary Tool nor EMBA reports it.

The binaries themselves, however, often identify their components: version banners, copyright notices, build paths, and library names are written for humans and survive stripping. The strings of \texttt{wcsinfo} state that the program is part of the Astrometry.net suite, give its release 0.93, and credit QFITS. Language models can read such text without a predefined vocabulary, which makes them a natural instrument for recovering components that no reference contains. Reading, however, is not recovering. The same strings contain the shared-library name \texttt{libgsl.so.27} and symbol versions such as \texttt{GLIBC\_2.35}, whose numbers are not releases, and the release 6.2.0 belongs to QFITS rather than to Astrometry.net. A model that reports every plausible name will present libraries that the binary merely loads as if they were embedded, attach releases to the wrong components, and map names to the wrong project; given these strings, the language model in our pipeline proposes cfitsio, a related astronomy library that no string of the file names. A natural safeguard is to accept only names that a product catalogue lists. It suppresses such errors, but it confines recovery to the catalogue's vocabulary, which lacks Astrometry.net. This paper therefore asks: \textbf{when a binary ships without an SBOM and its components appear in no reference, how much of its composition can be recovered from the binary alone, and how precisely?}

We answer this question with \TOOLNAME{}, an LLM-assisted tool that recovers SBOM records from ELF binaries without requiring their components to be in any reference. Its central idea is to separate proposing a component from accepting it: a language model proposes candidate components and versions from evidence retrieved out of the binary's strings, and deterministic checks against the binary's own structure and strings decide which candidates are reported. A vulnerability catalogue helps to propose names and to look up vulnerabilities. Each record states whether the component is part of the file or only a run-time dependency, and which evidence supports it~\cite{spdx23,ecma424cyclonedx16}. We compare \TOOLNAME{} with CVE Binary Tool, EMBA, Syft, and BLADE on 1,173 ELF files from Debian, drawn so that their components are in neither BLADE's feature database nor our catalogue, and additionally analyze 101 proprietary firmware images. Three findings stand out. First, reference-bound tools are nearly blind outside their references: BLADE identifies 3 of the 1,173 components and the scanners at most one, whereas \TOOLNAME{} identifies 464 (39.6\%), reports the correct release for 44.5\% of the components it recovers, and is correct for nearly four in five of the components it reports. Second, the binary sets the ceiling, and \TOOLNAME{} reaches far toward it: 198 of the 1,173 files never name their project and only 42.6\% of the version-labelled files carry their release in any string, but where a single string carries both the name and the release, \TOOLNAME{} reports the correct release for 89.3\% of the components it recovers. Third, the gain carries over to the field: on the firmware images, 148 of the 193 products that \TOOLNAME{} reports (77\%) appear in the output of neither CVE Binary Tool nor EMBA.

In summary, this paper makes the following contributions:
\begin{itemize}
    \item We propose \TOOLNAME{}, an LLM-assisted tool that recovers SBOM records from ELF binaries without requiring their components to be in any reference. An LLM proposes candidate components and versions from strings retrieved from the binary, in two passes, one guided by a vulnerability catalogue and one without it. A deterministic acceptance step then reports a candidate only when the binary itself provides evidence for it, in its ELF dynamic section, its strings, or its file name. Each record states whether the component is embedded in the binary or dynamically linked, and which evidence supports it.
    \item We construct, to our knowledge, the first benchmark whose targets are selected to lie outside the references that binary component identification tools rely on. It consists of 1,173 ELF files from 718 Debian projects across four architectures, whose components appear in neither the feature database of BLADE nor the product catalogue of CVE Binary Tool. We release the benchmark, \TOOLNAME{}, and all experimental data.
    \item On this benchmark, \TOOLNAME{} recovers the component of 39.6\% of the files with a precision of 78.7\%, whereas BLADE, CVE Binary Tool, EMBA, and Syft recover at most 0.3\%. Our ablation shows that the pass without the catalogue contributes most of this recall, and that the acceptance step keeps the output to about one record per file, instead of 6.47 without it. Analyzing a file costs less than half a US cent.
    \item We apply \TOOLNAME{} to 101 proprietary firmware images from network security appliances and automotive systems, where it reports 148 products that neither CVE Binary Tool nor EMBA reports.
\end{itemize}

The rest of the paper is organized as follows. Section~\ref{sec:background} introduces the SBOM content model, the evidence that survives in ELF binaries, and reference-based identification, and derives the challenges that \TOOLNAME{} addresses. Section~\ref{sec:methodology} presents \TOOLNAME{}. Section~\ref{sec:evaluation} describes the experimental setup, answers the four research questions, and discusses the implications of the results. Section~\ref{sec:threats} examines limitations and threats to validity, Section~\ref{sec:related} reviews related work, and Section~\ref{sec:conclusion} concludes.

\section{Background and Motivation}
\label{sec:background}

This section introduces the SBOM content model that binary recovery must populate, the evidence that survives in a deployed ELF binary, and the reference-based approaches that existing tools use to turn that evidence into component identities. It then derives the challenges and requirements that shape the design of \TOOLNAME{}.

\subsection{SBOMs and How Inventories Are Produced}
\label{subsec:bg-sbom}

Software Bill of Materials (SBOM) generation has become a foundational practice in software supply chain security, driven by the need for transparency and vulnerability traceability after major supply chain incidents, and by policy requirements that increasingly encourage or mandate SBOM adoption~\cite{14028,solarwind,eu_SBOM}. Automated SBOM and software composition analysis tools include Intel CVE Binary Tool~\cite{cvebintool}, EMBA~\cite{emba}, FACT~\cite{fact}, and Syft~\cite{syft}. Across this ecosystem, component information comes from two broad sources: metadata exposed by the build or packaging process, such as manifests, lockfiles, and package databases, and evidence retained in the artifact itself. For metadata-based generation, the central task is to extract the dependency information recorded by the build and map it into an interoperable representation~\cite{SBOM_lr}.

The two dominant SBOM standards, SPDX and CycloneDX, describe more than a flat list of names and versions~\cite{spdx,cyclonedx}. Both distinguish a component that is \emph{contained in} an artifact from one the artifact \emph{depends on} at run time. SPDX~2.3 defines relationship types such as \texttt{CONTAINS}, \texttt{DEPENDS\_ON}, \texttt{STATIC\_LINK}, and \texttt{DYNAMIC\_LINK}~\cite{spdx23}, and SPDX~3.0.1 retains this distinction through its \texttt{hasStaticLink} and \texttt{hasDynamicLink} relationships~\cite{spdx301}. CycloneDX~1.6 records dependencies in a separate graph and lets each component carry an \texttt{evidence} object~\cite{ecma424cyclonedx16}, whose \texttt{identity} entries state which field was inferred, with what confidence, and by which technique, including binary analysis, and whose \texttt{occurrences} point to the locations that support the claim. The standards therefore already provide a place for two kinds of information that binary-only recovery produces: how a component relates to the analyzed file, and what evidence supports its identity.

Metadata-based generation has an availability requirement: the artifact must still carry enough structured dependency information to reveal its composition. This premise weakens substantially after deployment, in stripped executables, repackaged firmware, proprietary embedded software, and vendor-specific binary bundles, where source code, debug symbols, manifests, and package databases may all be absent~\cite{empirical}. In these metadata-sparse settings, conventional tools may return incomplete or empty inventories, particularly for statically linked, vendored, or modified components~\cite{aaos_assess}. Component recovery must then rely on what the deployed binary itself retains.

\subsection{What Survives in a Deployed ELF Binary}
\label{subsec:bg-evidence}

A stripped ELF binary is not semantically empty. Two kinds of evidence typically survive stripping, and they differ in what they can prove.

\paragraph{Structural metadata.}
Some information is required by the ELF format or by the dynamic loader and is therefore preserved by \texttt{strip}. The dynamic section lists, in \texttt{DT\_NEEDED} entries, the shared objects that the file loads at run time, and a shared library declares its own name in \texttt{DT\_SONAME}~\cite{tis1995elf}. GNU symbol versioning additionally records, in the \texttt{.gnu.version\_r} section, the version nodes that the file requires from each needed library, such as \texttt{GLIBC\_2.14}~\cite{lsb50core}. These entries are exact and cheap to read, but they describe run-time dependencies rather than embedded code, and a version node is a lower bound on a compatible library, not the release that will be installed. A shared-library name such as \texttt{libffi.so.7} encodes an ABI version that need not match the upstream release that produced it. A few toolchains also embed package-level metadata: since Go~1.18, executables built by the \texttt{go} command embed module versions, version-control information, and build flags~\cite{go118notes}; Rust binaries built with \texttt{cargo-auditable} embed a compressed JSON list of their dependencies in a dedicated linker section~\cite{cargo_auditable}; and the UAPI package-metadata convention stores the package identity of a file as JSON in an ELF note~\cite{uapi8}. Where present, such metadata names embedded components exactly, but it is opt-in, language-specific, and absent from most C and C++ code. Statically linked C and C++ libraries leave no structural marker at all.

\paragraph{Residual text.}
The remaining evidence is textual: version banners, copyright and license notices, URLs, build and source paths, log and usage messages, and the name prefixes of exported or imported symbols. Earlier binary reuse detectors already exploit such residue: BAT compares string literals against package repositories~\cite{hemel2011bat}, and OSSPolice uses strings and exported function names when matching native binaries to open-source software~\cite{duan2017osspolice}. Residual text can reveal the identity of statically embedded code and of the project that built the file, which structural metadata cannot. It is also noisy. The same binary may mention libraries it merely links against, external tools it invokes, protocols it implements, and products it interoperates with, and many strings have no bearing on composition.

\paragraph{How often versions survive.}
Textual evidence does not guarantee that a release can be recovered. LibvDiff reports that only 42.5\% of 1,405 open-source library versions from the Conan package manager contain version strings~\cite{dong2024libvdiff}. A manual study of the 50 native libraries most often vendored into Python packages found that 24\% expose their version as a delimited, greppable string, 66\% encode it in a form that static string matching cannot discover, and 10\% embed no version information at all~\cite{alexopoulos2026pyxsieve}. Any recovery method that reads only the binary therefore faces an evidence ceiling: for a substantial share of components, the identity may be recoverable while the release is not, and an explicit \emph{unknown} version is a legitimate outcome rather than a failure to be hidden.

\subsection{Reference-Based Component Identification}
\label{subsec:bg-reference}

Automated identification of components in binaries compares what is observed in the binary with some reference that defines which identities can be reported. Existing approaches differ mainly in the kind of reference they use, and this choice bounds what they can find.

\paragraph{Name catalogues and signature rules.}
Scanner-oriented tools pair a list of product names with hand-written detection rules. CVE Binary Tool runs one checker per product, each consisting of patterns that match the product's banner and version strings~\cite{cvebintool}; EMBA applies a set of string-based version rules~\cite{emba}; and Syft's binary cataloger uses classifiers that match file names and version patterns for a fixed set of programs~\cite{syft}. The product vocabulary of vulnerability databases plays the same role at a larger scale: a recovered record is useful for vulnerability matching only if its name can be aligned with that vocabulary. Name catalogues are inexpensive to extend, because adding an identity does not require building any binary, but every detection rule must still be written by hand, and the catalogues themselves are inconsistent. A recent analysis of NVD and CVEdetails data found inconsistent vendor and product naming for a large share of entries, and found that NVD configurations predominantly model firmware as operating-system and hardware pairs rather than as application components~\cite{jiang2025vulcpe}.

\paragraph{Compiled-feature databases.}
Similarity-based detectors build their reference by compiling or indexing the source of known libraries and extracting features from the results, such as strings, symbols, learned function representations, and call graphs~\cite{tang2022libdb,yang2022modx,li2024libam,jiang2024binaryai,zou2025bincofer,zhao2024firmsec,dong2024libvdiff} (Section~\ref{subsec:rw-sca}). BLADE retrieves candidate libraries by matching strings and symbols against a corpus of several thousand libraries and then asks an LLM to accept or reject each candidate using structured identity, metadata, and overlap evidence~\cite{liu2026blade}. These systems achieve high precision when the target library is in the reference, and some can recover releases that leave no version string~\cite{dong2024libvdiff}.

\paragraph{The reference as a recall bound.}
In both families, a component can be reported only if the reference already contains it. The literature acknowledges this bound but rarely measures it. LibDB can report only reuses that are present in its database and falls back to the closest known version when the exact target is missing~\cite{tang2022libdb}, and FirmSec states that components absent from its database become false negatives~\cite{zhao2024firmsec}. Evaluations, moreover, are often constructed so that the bound does not apply. BLADE's reference corpus was built to cover every library in its test set, and part of its false positives arise when the true library is not among the retrieved candidates and a related library is reported instead~\cite{liu2026blade}. A recent study of eleven tools likewise built the databases of the open-source tools to contain every test library; it suggests that the low recall of the commercial tools stems partly from limited database coverage, but because their databases are closed, which test libraries they lack is unknown~\cite{liu2026binarytpl}. We are not aware of a published evaluation that reports recall for components outside the tool's own reference. The two kinds of reference also differ in cost and breadth. A compiled-feature database requires that someone has built each library, across versions and architectures, before it can be detected; a name catalogue requires only that the identity has been named. Neither covers vendor-only components that were never compiled into a public corpus and never registered in a vulnerability database, which are common in proprietary firmware.

\paragraph{LLMs inside a reference.}
Language models are well suited to interpreting residual text, because banners, paths, and messages are written for humans. Recent detectors use them, but keep them inside a reference: the model selects among retrieved candidates or writes rules for libraries whose versions are known (Section~\ref{subsec:rw-llm}). Without a reference to constrain them, however, language models can also produce plausible but unsupported claims. Methods that revise generated text against retrieved evidence~\cite{gao2023rarr} or separate generation from verification~\cite{dhuliawala2024cove} reduce this risk, but intrinsic self-correction without external feedback is unreliable~\cite{huang2024selfcorrect}. For binary recovery, the external feedback must come from the binary itself.

\subsection{Challenges}
\label{subsec:bg-challenges}

Figure~\ref{fig:demo} illustrates the problem on a single binary that contains names such as \texttt{BoringSSL} and \texttt{Netscape} alongside release-like numbers and protocol identifiers such as \texttt{TLSv1.2}. Recovering scanner-facing records from such evidence involves five challenges.

\begin{itemize}
    \item \textbf{Sparse and selective evidence.} Identifying strings are few, unevenly distributed, and absent for some components. A version may be missing even when the identity is visible.
    \item \textbf{Presence versus reference.} A name in a binary can denote statically embedded code, a dynamically linked library, an external program invoked at run time, or a mere mention. For example, the string \texttt{libcairo.so.2} in an application plugin indicates a run-time dependency, and \texttt{GLIBC\_2.14} indicates a version requirement, not embedded C library code.
    \item \textbf{Identity normalization.} Observed names rarely coincide with canonical identities. Sub-artifacts carry their own names, such as the \texttt{libisc} and \texttt{libisccfg} libraries shipped by BIND; similar names can belong to unrelated projects, such as glibc's \texttt{libnss\_compat} and Mozilla NSS; and generic words such as \texttt{agent} or \texttt{printer} collide with registered product names.
    \item \textbf{Version binding.} A binary often contains several version-like tokens, including library ABI numbers, protocol versions, build identifiers, and the versions of other components. Associating the correct token with the correct component is distinct from finding a token~\cite{dong2024libvdiff}.
    \item \textbf{Out-of-reference components.} Components that are neither compiled into a feature corpus nor named in a catalogue cannot be reported by reference-bound methods, yet they may still identify themselves through banners, paths, or library names.
\end{itemize}

\section{Methodology}
\label{sec:methodology}

\begin{figure*}[ht]
    \centering
  \includegraphics[width=\linewidth]{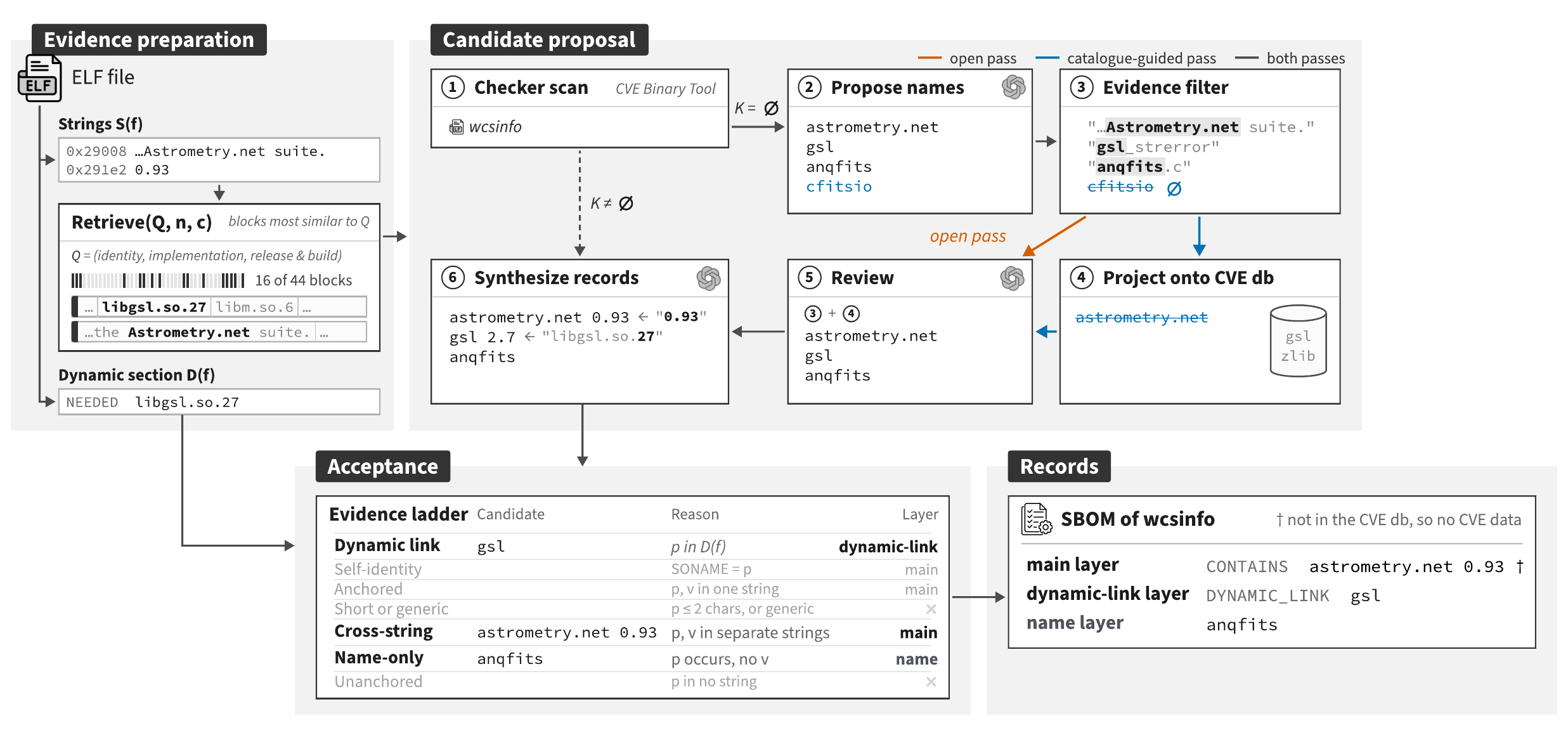}
  \caption{Overview of \TOOLNAME{}{} on the example \texttt{wcsinfo}.}
  \label{fig:arch}
\end{figure*}

\subsection{Problem Formulation and Design Goals}
\label{subsec:formulation}

Given an ELF file $f$, an executable or a shared object that may have been extracted from a firmware image, \TOOLNAME{} recovers the software components that $f$ contains or depends on. Package metadata for $f$ is assumed to be unavailable, and its symbol tables may have been stripped. PE files are out of scope: the structural evidence described in Section~\ref{subsec:ladder} is specific to ELF, and every corpus in our evaluation consists of ELF files.

Let the catalogue $P$ denote the finite vocabulary of product names used by a vulnerability scanner; in our implementation, $P$ consists of the distinct product names in the CVE range data of CVE Binary Tool~\cite{cvebintool}. For each $p \in P$, let $V(p)$ denote a short text summary of the version examples and range endpoints recorded for $p$. The output for a single file is a set of records
\[
O_f = O_f^{\mathrm{main}} \cup O_f^{\mathrm{dyn}} \cup O_f^{\mathrm{name}}, \qquad r_i = (p_i, v_i, \rho_i, \tau_i, E_i, \kappa_i),
\]
where $p_i$ is a product name, which need not belong to $P$; $v_i$ is a version string, or $\bot$ (undefined) when no version is supported; $\rho_i \in \{\text{self}, \text{embedded}, \text{dynamic-link}\}$ states whether the record is the identity that the file declares in its \texttt{DT\_SONAME}, a component whose code the file contains, or a library that the file loads at run time; $\tau_i$ is the evidence tier of the record; $E_i$ is the supporting evidence, namely the strings and the entries of the ELF dynamic segment that place the record in its tier; and $\kappa_i \in P \cup \{\bot\}$ is the catalogue product used for vulnerability lookup, or $\bot$ when no catalogue product corresponds. The acceptance step admits records on the evidence of the file and places them in three layers. The main layer $O_f^{\mathrm{main}}$ holds the self and embedded records, that is, the components that the file declares or contains. The dynamic-link layer $O_f^{\mathrm{dyn}}$ holds the libraries that the file loads at run time. The name layer $O_f^{\mathrm{name}}$ holds components that the binary merely names, without evidence for a version and without a match to the file's own name; they are reported apart, with $v_i = \bot$. A single file can contain or load several components, such as the project that built it, statically linked libraries, and shared libraries loaded at run time, so $O_f$ is a set rather than a single record. An empty set expresses abstention.

\TOOLNAME{} separates proposing a record from accepting it. Two passes of the same staged LLM pipeline propose candidate components and versions from evidence that is retrieved, stage by stage, out of the complete string content of $f$. The catalogue-guided pass maps the proposed names onto $P$; the open pass keeps them as proposed, so that a component absent from $P$ can still become a candidate. Both passes show the model the version hints $V(p)$ of candidates that $P$ contains. The open pass reaches components that the catalogue lacks, whereas the guided pass supplies the catalogue spellings and version hints for the components that the catalogue lists (Section~\ref{subsec:rq2}). A deterministic acceptance step then examines the candidates of both passes alike. Consulting only the ELF dynamic segment, the printable strings, and the name of $f$, together with one statistic over the files analyzed together, it places each candidate in the main, dynamic-link, or name layer, or drops it, and assigns its relationship and tier. The dynamic segment also yields a record directly: a file that declares its own name in \texttt{DT\_SONAME} receives a self record for that name. Figure~\ref{fig:arch} gives an overview and follows one executable through it, \texttt{wcsinfo} of the Debian project Astrometry.net. Blue marks the catalogue-guided pass, vermillion the open pass, and black what both passes share; a logo marks a model call, a struck-out name is dropped, and $\emptyset$ marks an empty result. For brevity, the figure omits two components that \TOOLNAME{} reports for this file, qfits and glibc. Algorithm~\ref{alg:relicbom} at the end of Section~\ref{subsec:ladder} summarizes the complete procedure. Section~\ref{subsec:evidence} describes the evidence extracted from $f$, Section~\ref{subsec:proposal} the proposal stages, and Section~\ref{subsec:ladder} the acceptance step and the catalogue annotation.

The design pursues five goals.
\begin{itemize}
    \item \textbf{Evidence-decided acceptance.} Every record is accepted by evidence from the file and carries that evidence $E_i$, so that each record can be checked against the binary and each acceptance decision can be reproduced.
    \item \textbf{Relationship awareness.} The relationship $\rho_i$ separates the identity that the file declares, embedded components, and run-time dependencies. Run-time dependencies are reported in the dynamic-link layer, and components that the binary merely names are kept apart in the name layer, so that the main layer lists only what the file declares or contains.
    \item \textbf{Catalogue as proposal and annotation.} The catalogue may propose names and version hints, and it supplies $\kappa_i$ for vulnerability lookup. Membership in $P$ is never a condition for reporting a component, and a catalogue match alone never suffices.
    \item \textbf{Honest versions.} A version is reported only when the binary supports it; otherwise $v_i = \bot$. Implausible version tokens are discarded rather than guessed, and numbers that the binary shows only as an ABI version or a symbol-version requirement are not reported as releases.
    \item \textbf{Error isolation.} Recovery is decomposed into stages, so that an early error, such as a spurious candidate name, must survive later checks before it can become a record.
\end{itemize}

Two boundaries delimit the scope. First, recovery is performed per file: apart from one statistic over the files analyzed together (Section~\ref{subsec:ladder}), the records of a file depend on that file alone. The records of several files can be merged into a summary of an image, but \TOOLNAME{} does not infer dependencies across files or compute a transitive closure. Second, \TOOLNAME{} relies on textual evidence and ELF structure only, and performs no function-level similarity matching against compiled references. A component that leaves neither identifying text nor a dynamic-linking entry in $f$ is therefore outside its reach.

\subsection{Evidence Preparation}
\label{subsec:evidence}

Before any model call, \TOOLNAME{} extracts two kinds of evidence from $f$: the entries of its ELF dynamic segment and its printable strings. Both are computed once per file, shared by every later stage and by both passes, and obtained without the language model, the catalogue, or version parsing.

\paragraph{Structural evidence.}
Let $X_f$ denote the byte sequence of $f$. A parser reads the ELF header and the program headers of $X_f$, maps virtual addresses to file offsets through the loadable segments, and decodes the dynamic segment. It needs no section headers and therefore also works on stripped files. It yields
\[
D(f) = \bigl(N(f),\; s(f),\; \Phi(f)\bigr),
\]
where $N(f)$ is the list of \texttt{DT\_NEEDED} entries, that is, the shared libraries that $f$ loads at run time; $s(f)$ is the \texttt{DT\_SONAME} entry, the name under which a shared library declares itself, or $\bot$ if there is none; and $\Phi(f)$ maps each needed library to the symbol versions that $f$ requires from it, such as \texttt{GLIBC\_2.34} from \texttt{libc.so.6}. A statically linked executable has no dynamic segment, so that $N(f)$ and $\Phi(f)$ are empty and $s(f) = \bot$.

\paragraph{Textual evidence.}
The textual evidence of $f$ is the set
\[
S(f) = \{(o_j, w_j)\}_{j=1}^{|S(f)|}
\]
of all maximal runs $w_j$ of at least four printable ASCII characters or tabs in $X_f$, the default minimum of the \texttt{strings} utility, each with its byte offset $o_j$. We call these runs the strings of $f$. No run is filtered out: copyright notices, usage messages, symbol names, and paths are kept alongside version strings, because any of them can identify a component.

\paragraph{Evidence store.}
To let each stage read the part of $S(f)$ that concerns it, the runs are organized into an evidence store. Runs longer than a block length $L$ are split into pieces of at most $L$ characters, and consecutive pieces are packed, in file order, into blocks of at most $L$ characters. Each distinct block $b$ is embedded once by a text-embedding model into a unit vector $\mathbf{e}_b$, while all of its occurrences and offsets remain in the store. The store is bound to the digest of $f$ and built only once.

\paragraph{Retrieval.}
A stage obtains evidence through an operator $\textsc{Retrieve}(Q, n, c)$ that takes a list of natural-language queries $Q = (q_1, \ldots, q_{|Q|})$, each embedded by the same model into a unit vector $\mathbf{e}_q$, a maximum number of blocks $n$, and a character budget $c$. It selects blocks greedily by maximal marginal relevance~\cite{carbonell1998mmr}, with the queries taking turns. At turn $t = 0, 1, \ldots$, let $B_t$ be the blocks selected so far and $A_t$ the blocks not yet considered; the operator selects
\[
b_t = \operatorname*{arg\,max}_{b \in A_t} \Bigl[\, \lambda\, \mathbf{e}_b^{\top} \mathbf{e}_{q_{(t \bmod |Q|)+1}} \;-\; (1-\lambda) \max_{b' \in B_t} \mathbf{e}_b^{\top} \mathbf{e}_{b'} \Bigr],
\]
where the maximum over an empty set is $0$ and $\lambda \in [0, 1]$ weighs relevance against redundancy. A block that would exceed the remaining budget is skipped and not considered again, and selection stops after $n$ blocks or when no block fits. The alternation of queries keeps a single query, for example one of several candidate components, from consuming the budget, and the redundancy term keeps near-duplicate blocks, such as repeated symbol names, from filling the context. The proposal stage of each pass retrieves with the budget $(n_1, c_1)$, and the stages that examine specific candidates retrieve with $(n_2, c_2)$ (Section~\ref{subsec:proposal}). Retrieval accepts or rejects nothing by itself, but every later stage sees only the blocks it returns, which is why each stage retrieves for its own candidates.

Figure~\ref{fig:evidence-example} shows both kinds of evidence for \texttt{wcsinfo}, an executable of the Debian project Astrometry.net, to which Section~\ref{subsec:rq1} returns. Its strings fill 44 blocks, of which the proposal stage retrieves 16 (Figure~\ref{fig:arch}, top left).

\begin{figure}[t]
\centering
\setlength{\fboxsep}{4pt}
\fcolorbox{black!35}{black!5}{\begin{minipage}{0.97\linewidth}\centering\small
\texttt{/usr/bin/wcsinfo}\quad Debian project Astrometry.net, release 0.93; amd64, 232{,}960 bytes
\end{minipage}}\\[1pt]
{\large$\Downarrow$}\\[1pt]
\fcolorbox{black!35}{white}{\begin{minipage}[t][3.3cm][t]{0.44\linewidth}
\textbf{\small Structural evidence} $D(f)$\par\smallskip
\scriptsize
\begin{tabular}{@{}l@{\hspace{5pt}}l@{}}
$s(f)$ & $\bot$ (an executable declares no name) \\[2pt]
$N(f)$ & \texttt{libgsl.so.27} \\
       & \texttt{libm.so.6} \\
       & \texttt{libc.so.6} \\[2pt]
$\Phi(f)$ & \texttt{libc.so.6}: \texttt{GLIBC\_2.34}, \texttt{GLIBC\_2.33}, \ldots \\
       & \texttt{libm.so.6}: \texttt{GLIBC\_2.35}, \texttt{GLIBC\_2.29}, \ldots
\end{tabular}
\end{minipage}}\hfill
\fcolorbox{black!35}{white}{\begin{minipage}[t][3.3cm][t]{0.49\linewidth}
\textbf{\small Textual evidence} $S(f)$\par\smallskip
\scriptsize
\begin{tabular}{@{}l@{\hspace{5pt}}l@{}}
\multicolumn{2}{@{}l}{2{,}251 runs, packed into 44 blocks} \\[2pt]
\textit{offset} & \textit{run} \\
\texttt{0x29008} & \texttt{This program is part of the}\ldots \\
                 & \quad\ldots\texttt{Astrometry.net suite.} \\
\texttt{0x291e2} & \texttt{0.93} \\
\texttt{0x2c0d8} & \texttt{Written by QFITS} \\
\texttt{0x2c6bc} & \texttt{6.2.0} \\
                 & \ldots\ 32 runs contain \texttt{qfits}
\end{tabular}
\end{minipage}}\\[1pt]
{\large$\Downarrow$}\\[1pt]
\fcolorbox{black!35}{black!5}{\begin{minipage}{0.97\linewidth}\footnotesize
\textcolor{green!45!black}{$\checkmark$}\ \textbf{Identity} Astrometry.net: named by four runs, such as the banner at \texttt{0x29008}.\par
\textcolor{green!45!black}{$\checkmark$}\ \textbf{Release} 0.93: present, but in a run of its own, apart from the name.\par
\textcolor{green!45!black}{$\checkmark$}\ \textbf{Bundled library} qfits: named by 32 runs, with its release 6.2.0 in a separate run.\par
\textcolor{green!45!black}{$\checkmark$}\ \textbf{Dependency evidence}: the three libraries of $N(f)$.\par
\textcolor{red!70!black}{$\times$}\ \textbf{Not releases}: the \texttt{27} in \texttt{libgsl.so.27} is an interface version, and \texttt{GLIBC\_2.35} a symbol version.
\end{minipage}}
\caption{Evidence that \TOOLNAME{} extracts from a Debian executable before any model call. Left: the structural evidence $D(f)$ from the ELF dynamic segment; the executable declares no name of its own, loads three libraries, and requires symbol versions from two of them. Right: four of its printable runs $S(f)$ with their byte offsets. Bottom: the runs name the project and its release 0.93, though in separate strings, as well as qfits, a library that Astrometry.net bundles, with its release 6.2.0. The numbers in \texttt{libgsl.so.27} and in the symbol versions are interface versions, not releases.}
\label{fig:evidence-example}
\end{figure}

\subsection{Candidate Proposal}
\label{subsec:proposal}

Both passes run the same staged pipeline over the evidence of Section~\ref{subsec:evidence}. They differ in three points, all of them in the catalogue-guided pass: it adds names from package labels to the proposal, projects the names onto the catalogue, and shows the version hints $V(p)$ in the review. Each stage retrieves the blocks that answer its own question, and deterministic checks between the language-model calls discard candidates that the strings do not support. A pass returns candidate records only; which of them are reported is decided in Section~\ref{subsec:ladder}. Table~\ref{tab:stages} lists the stages and traces both passes on one library, Figure~\ref{fig:arch} follows them on \texttt{wcsinfo}, Algorithm~\ref{alg:pass} gives the control flow of a pass, including its early exits, and Figure~\ref{fig:prompts} shows excerpts of the three prompts. We write $\textsc{LLM}(\pi, T)$, or $\textsc{LLM}(\pi, T, X)$ when the prompt also lists a candidate set $X$, for a model call with prompt $\pi$ on retrieved blocks $T$, whose payload also contains the file name and the number of blocks, and $\psi$ for a parser that extracts the first well-formed JSON object of the reply.

\begin{table*}[t]
\centering
\caption{The stages of one proposal pass, read from top to bottom; each stage hands its output to the next. \TOOLNAME{} runs this pass twice per file: the catalogue-guided pass maps the names onto catalogue entries in stage~4, and the open pass skips that stage. The two right-hand columns trace both passes on \texttt{libmsgpack-c.so.2.0.0}, a Conan build of msgpack-c~6.0.0, a project that the catalogue does not list. Both passes propose the same names, but the projection drops msgpack-c and adds the look-alike entry gpac, so that only the open pass yields the target with its release. The glibc versions are symbol-version requirements (\texttt{GLIBC\_2.x}); the acceptance step reports glibc as a run-time dependency (Section~\ref{subsec:ladder}).}
\label{tab:stages}
\small
\setlength{\tabcolsep}{4pt}
\begin{tabular}{@{}l>{\raggedright\arraybackslash}p{0.11\textwidth}>{\raggedright\arraybackslash}p{0.28\textwidth}c>{\raggedright\arraybackslash}p{0.19\textwidth}>{\raggedright\arraybackslash}p{0.19\textwidth}@{}}
\toprule
& & & & \multicolumn{2}{c}{Example: \texttt{libmsgpack-c.so.2.0.0}} \\
\cmidrule(l){5-6}
& Stage & Operation & Output & Catalogue-guided pass & Open pass \\
\midrule
1 & Checker scan & Run the version checkers of CVE Binary Tool. & $K$ & none & none \\[3pt]
2 & Proposal & LLM ($\pi_1$) proposes component names, without versions. & $H$ & msgpack-c, glibc & msgpack-c, glibc \\[3pt]
3 & Evidence filter & Keep names with evidence score $\gamma(p) \ge \theta_1(p)$. & $H'$ & msgpack-c, glibc & msgpack-c, glibc \\[3pt]
4 & Projection & Map names to catalogue entries; drop unmatched names. & $C$ & glibc, gpac & \textit{skipped:} msgpack-c, glibc \\[3pt]
5 & Review & LLM ($\pi_2$) drops unsupported candidates; keep those with $\gamma(p) \ge \theta_2(p)$. & $C'$ & glibc & msgpack-c, glibc \\[3pt]
6 & Synthesis & LLM ($\pi_3$) writes product--version records; clean versions, merge $K$. & $R$ & glibc 2.3, glibc 2.14, glibc 2.2.5 & \textbf{msgpack-c 6.0.0}, glibc 2.14 \\
\bottomrule
\end{tabular}
\end{table*}

\paragraph{Checker scan.}
Before the first model call, the version checkers of CVE Binary Tool~\cite{cvebintool}, hand-written signatures that pair a catalogue product with a version pattern, are run on $f$; hits for the compiler are dropped. The resulting pairs $K$ need no model: their products seed the proposal, they are exempt from the evidence filters below, and they are merged into the records of the pass. In Figure~\ref{fig:arch}, the dashed arrow carries $K$ to the records; \texttt{wcsinfo} has no hits.

\paragraph{Proposal.}
The proposal stage retrieves $T_1 =\textsc{Retrieve}(Q_{\mathrm{disc}}, n_1, c_1)$, where $Q_{\mathrm{disc}}$ consists of three fixed queries that ask for the identity of the software, its implementation, and its release and build information. If $T_1$ is empty, the pass returns $K$ without a model call. Otherwise the model proposes names of products, libraries, or operating-system modules, but no versions:
\[
H = \psi\bigl(\textsc{LLM}(\pi_1, T_1)\bigr).
\]
If the model proposes nothing, a small set of token rules for unmistakable banners, such as \texttt{ffmpeg version}, supplies names instead. The products of $K$ are added to $H$; in the catalogue-guided pass, so are names that occur in package labels of the form \texttt{version=}\textit{name}. Such a label may name third-party software and can therefore propose a catalogue candidate, but it cannot support one by itself, which is why the evidence score below penalizes a name whose only occurrence is a label. Abbreviations in parentheses are split off, and a fixed alias table adds the catalogue spelling of names that are known to differ from it, such as \texttt{wpa\_supplicant} for \texttt{wpa\_cli}.

\paragraph{Evidence filter.}
Every proposed name must then be supported by the strings. The stage retrieves $T_2 = \textsc{Retrieve}(Q_{\mathrm{id}}(H), n_2, c_2)$ with one query per name, which asks for identity and implementation evidence and for the distinction between included software and external mentions, and computes a deterministic evidence score $\gamma(p)$ over $T_2$. The score adds points when the name occurs as a whole token, in the form \texttt{lib}\textit{name}, or next to a banner pattern recorded for the product, and it subtracts points when the only occurrence of the name is a package label, when the name is a generic catalogue entry that collides with ordinary words, or when the strings or the file name mark a vendor-internal module. The banner patterns and the markers of vendor-internal modules were collected while we developed \TOOLNAME{} on proprietary automotive firmware, and they are part of the released artifact. The score only filters proposals; the acceptance step of Section~\ref{subsec:ladder} uses none of these rules. A name is kept if
\[
\gamma(p) \ge \theta_1(p) \quad\text{or}\quad p \text{ stems from } K,
\]
where $\theta_1(p)$ takes one of three values: a general one, a higher one for generic names that collide with catalogue entries, and the highest for platform and API names such as OpenGL~ES. The kept names form $H'$; if no name remains, the pass returns $K$. On \texttt{wcsinfo}, the catalogue-guided pass proposes cfitsio, which no string names, and the filter removes it.

\paragraph{Projection.}
In the catalogue-guided pass, each remaining name is mapped onto the catalogue: an exact case-insensitive match is tried first, then a match of normalized core names, then a whole-word match against catalogue names of similar length, and finally a fuzzy match with a similarity ratio of at least $r_{\min}$. The matches are ranked by how closely they follow the proposed names, and the best $k_P$ form the candidate set $C \subseteq P$. The open pass skips this stage and keeps the proposed names, $C = H'$, so that a component absent from the catalogue can still become a candidate. On \texttt{wcsinfo}, only Astrometry.net reaches this stage in the catalogue-guided pass, and the catalogue has no entry for it.

\paragraph{Review.}
The stage retrieves $T_3 = \textsc{Retrieve}(Q_{\mathrm{id}}(C), n_2, c_2)$, and the model removes the candidates that $T_3$ does not support:
\[
C' = \psi\bigl(\textsc{LLM}(\pi_2, T_3, C)\bigr) \cap C .
\]
The prompt lists the candidates and the proposed names they were derived from, because a catalogue spelling often differs from the strings; in the catalogue-guided pass, it also shows the version hints $V(p)$ of each candidate. If the reply cannot be parsed, all candidates are kept.

A second deterministic gate then requires $\gamma(p) \ge \theta_2(p)$, with the score now computed over $T_3$, unless $p$ stems from $K$; $\theta_2(p)$ uses the same three classes of names as $\theta_1(p)$, with a higher general value. The two checks fail in different ways: the model can tell an included component from a mere mention, and is instructed to keep catalogue spellings that differ from the strings, whereas the score cannot read context but rejects names that the strings barely support. The survivors are ranked by $\gamma(p)$. If no candidate survives but some proposed name in $H$ scores at least $\theta_{\mathrm{f}}$ and is not a vendor-internal module name, synthesis retrieves for the names in $H$ and runs without a candidate list; otherwise the pass returns $K$.

\paragraph{Synthesis.}
The last stage retrieves $T_4 = \textsc{Retrieve}(Q_{\mathrm{ver}}(C'), n_2, c_2)$ with one query per candidate, which asks for its release version and build identity together with the context that indicates whose version it is. The model then writes the records:
\[
R^{\circ} = \psi\bigl(\textsc{LLM}(\pi_3, T_4, C')\bigr), \qquad R = \sigma\bigl(R^{\circ}, K\bigr).
\]
The prompt restricts the products to $C'$, shows $V(p)$ for every candidate that the catalogue contains, and allows several records when distinct products, or distinct versions with clear evidence, are present. A record whose version cannot be associated with its product is kept with an empty version rather than dropped. The sanitizer $\sigma$ removes placeholder and platform names, clears implausible versions such as a single character or a major-only number for a product with dotted releases, merges the hits $K$, and, when a product occurs both with and without a version, keeps only the versioned records. All three prompts also instruct the model to abstain on package labels and vendor-internal module names, a rule from the same development on automotive firmware.

The two passes thus yield the candidate records $R_{\mathrm{guided}}$ and $R_{\mathrm{open}}$ of $f$.

\begin{algorithm}[t]
\caption{One proposal pass over a file $f$}
\label{alg:pass}
\begin{algorithmic}[1]
\Require file $f$ with its evidence store; mode $\in \{\text{guided}, \text{open}\}$
\Ensure candidate records $R_{\mathrm{mode}}$ of $f$
\State $K \gets$ checker hits on $f$; \quad $T_1 \gets \textsc{Retrieve}(Q_{\mathrm{disc}}, n_1, c_1)$
\State \textbf{if} $T_1 = \emptyset$ \textbf{then return} $K$
\State $H \gets \psi(\textsc{LLM}(\pi_1, T_1))$, or banner token rules if the reply is empty
\State add to $H$ the products of $K$, the aliases, and, if guided, the label names
\State $H' \gets \{p \in H : \gamma(p) \ge \theta_1(p) \text{ over } T_2\} \cup \text{products of } K$
\State \textbf{if} $H' = \emptyset$ \textbf{then return} $K$
\State $C \gets$ the best $k_P$ catalogue matches of $H'$ if guided, else $H'$
\State $C' \gets \psi(\textsc{LLM}(\pi_2, T_3, C)) \cap C$; keep $p$ with $\gamma(p) \ge \theta_2(p)$ over $T_3$ or $p$ from $K$
\State \textbf{if} $C' = \emptyset$ and no non-internal $p \in H$ has $\gamma(p) \ge \theta_{\mathrm{f}}$ \textbf{then return} $K$
\State $R^{\circ} \gets \psi(\textsc{LLM}(\pi_3, T_4, C'))$ \Comment{if $C' = \emptyset$: no list, $T_4$ for $H$}
\State \Return $\sigma(R^{\circ}, K)$
\end{algorithmic}
\end{algorithm}

\begin{figure*}[t]
\centering
\small
\renewcommand{\arraystretch}{1.25}
\begin{tabular}{|>{\raggedright\arraybackslash}p{0.30\textwidth}|>{\raggedright\arraybackslash}p{0.30\textwidth}|>{\raggedright\arraybackslash}p{0.30\textwidth}|}
\hline
\multicolumn{1}{|c|}{\textbf{Proposal ($\pi_1$)}} & \multicolumn{1}{c|}{\textbf{Review ($\pi_2$)}} & \multicolumn{1}{c|}{\textbf{Synthesis ($\pi_3$)}} \\
\hline
Based on the extracted strings, suggest potential product names, library names, or operating system module names that might be present in this binary. \ldots{} Suggest only product/library names (not versions). Ignore protocol related strings such as HTTP, HTTPS, IEEE, etc. Skip strings related to compiler supports such as GLIBC, GCC, CLANG, GLIBCXX etc. \ldots{} Be conservative -- only suggest names you are confident about. \ldots{}
&
Remove only the candidates that have NO supporting evidence in the extracted strings. \ldots{} Candidates that are a plausible canonical name for a product that IS evidenced in the strings must be KEPT, even if the exact spelling differs from the strings. \ldots{} When uncertain about OEM/internal components or generic catalogue collisions, REMOVE the candidate -- abstention is correct. \ldots{}
&
Create a SBoM (Software Bill of Materials) entry based on the string values. Identify the product names, library names or operating system module names. Also identify the respective version numbers. \ldots{} You MAY return MULTIPLE records when (a) distinct products are clearly present, or (b) the SAME product has multiple distinct, well-evidenced version strings \ldots{} Do NOT borrow a version number that clearly belongs to a DIFFERENT component \ldots{} \\
\hline
\texttt{\{"suggested\_products": [...]\}} & \texttt{\{"appropriate\_products": [...]\}} & \texttt{\{"explaination": "...", "records": [\{"product": "...", "version": "..."\}]\}} \\
\hline
\end{tabular}
\caption{Excerpts of the three prompts (``\ldots'' marks omitted text; the full prompts are part of the released artifact). The review and synthesis prompts also list the candidates, with their catalogue version hints $V(p)$ in the review of the catalogue-guided pass and, in both passes, in the synthesis of every candidate that the catalogue contains.}
\label{fig:prompts}
\end{figure*}

\subsection{Evidence-Gated Acceptance}
\label{subsec:ladder}

The acceptance step decides which candidate records are reported, with which relationship, and with which version. It pools the candidate records of both passes, $R_{\mathrm{guided}} \cup R_{\mathrm{open}}$, and treats each of them alike, whichever pass proposed it and whether or not its product belongs to $P$. Its decisions consult only the structural evidence $D(f)$, the strings of $f$, the file name of $f$, and one statistic over the batch; they involve no model call, no evaluation label, no catalogue lookup, and no product-specific rule beyond two alias tables, so that each decision can be reproduced without the model.

\paragraph{Occurrence and matching tests.}
The tests operate on the strings of $f$, not on blocks. A name $p$ \emph{occurs} in a string if the string contains $p$, or a variant of $p$ that swaps hyphens, underscores, and spaces or adds or removes a \texttt{lib} prefix, and the match is neither preceded by a letter or digit nor followed by a letter. A version $v$ occurs in a string if the string contains $v$ verbatim, with or without a leading \texttt{v}. Compiler and linker identification strings, such as \texttt{GCC: (GNU) 9.3.0}, describe the build environment rather than a component and are ignored by both tests. A product \emph{matches} a shared-library name, such as an entry of $N(f)$ or $s(f)$, if the product, with or without a \texttt{lib} prefix, equals a leading run of the name's tokens once the \texttt{lib} prefix and the \texttt{.so} suffix are removed. It matches a symbol-version requirement if the prefix of the version name, such as \texttt{ZLIB} in \texttt{ZLIB\_1.2.9}, equals the product or, for products of at least four characters, begins with it. Two alias tables, with entries for 121 and 70 products, cover products whose library or version names differ from the product name, such as glibc with \texttt{libc.so.6} and \texttt{GLIBC\_2.14}; both tables are part of the released artifact.

\paragraph{Evidence ladder.}
\begin{table*}[t]
\centering
\caption{Tiers of the evidence ladder for a candidate $(p, v)$ of file $f$, in the order in which they are checked; the first tier whose condition holds decides.}
\label{tab:ladder}
\small
\begin{tabular}{@{}>{\raggedright\arraybackslash}p{0.13\textwidth}>{\raggedright\arraybackslash}p{0.55\textwidth}>{\raggedright\arraybackslash}p{0.25\textwidth}@{}}
\toprule
Tier & Condition & Reported as \\
\midrule
Dynamic link & An entry of $N(f)$ or a symbol-version requirement in $\Phi(f)$ matches $p$. & dynamic-link record, dynamic-link layer \\[2pt]
Self-identity & $s(f)$ matches $p$. & self record, main layer \\[2pt]
Anchored & One string contains both $p$ and $v$; or $v = \bot$ and $p$ occurs in a versioned path segment, such as \texttt{/zlib-1.2.11/}, or in a shared-library name with a release-like version of at least two components, such as \texttt{libcivetweb.so.1.15.0}. & embedded record, main layer \\[2pt]
Short or generic & $p$ has at most two characters; or $v = \bot$ and $p$ is generic, that is, proposed without a version in at least $m_{\mathrm{g}}$ files of the batch and placed in one of the first three tiers in none of them. & not reported \\[2pt]
Cross-string & $p$ and $v$ occur, but never in the same string. & embedded record, main layer \\[2pt]
Name-only & $p$ occurs; $v$ occurs in no string, or $v = \bot$. & embedded record; main layer if $p$ is consistent with the identity of $f$, name layer otherwise \\[2pt]
Unanchored & $p$ occurs in no string. & not reported \\
\bottomrule
\end{tabular}
\end{table*}
The ladder of Table~\ref{tab:ladder} assigns each candidate $(p, v)$ the first tier, read from top to bottom, whose condition $f$ satisfies. The tier $\tau$ fixes the relationship $\rho(\tau)$ and the layer of the record, and the strings and dynamic entries that satisfy its condition become the evidence $E$ of the record. 
The order of the tiers follows how reliably each kind of evidence binds a version to a name. A name and a version printed in the same string, as in a banner, rarely belong to different components, and a library that declares its name in $s(f)$ identifies itself. A version that occurs only elsewhere in the binary may belong to another component. Such cross-string records are nevertheless reported, because many programs print their name and their version in separate strings, for example in a usage line and a version line; their tier tells the consumer that the binding is weaker. A shared-library name does not anchor a version when its trailing number is an ABI version, as in \texttt{libm.so.6}. Names of one or two characters, such as \texttt{go} or \texttt{qt}, collide with ordinary text and are therefore accepted only through a structural match or an anchor, such as \texttt{go1.11.6}. The generic-name condition is the only one computed across files: it uses the candidates of the batch, that is, of all files analyzed together, such as the ELF files of one firmware image, and removes names that the model proposes in many files without ever finding an anchor. Acceptance therefore runs once both passes have finished for every file of the batch; for a single file, the condition never applies. In Figure~\ref{fig:arch}, gsl matches \texttt{libgsl.so.27} in $N(f)$ and becomes a dynamic-link record; astrometry.net and its release 0.93 occur only in separate strings, which places the record in the cross-string tier; and anqfits, named without a version and unrelated to the file name, enters the name layer.

\paragraph{Identity consistency.}
A name-only record lacks version evidence, but it can still identify the file itself. This is common for executables, which have no $s(f)$ and often name their project in messages and paths without printing its release. A name-only record therefore enters the main layer if its name is consistent with the identity of $f$. The identity forms of $f$ are the stem of $s(f)$, the file name up to its first dot, and the whole file name, each reduced to lower-case letters and digits and taken with and without a leading \texttt{lib}. The name $p$, reduced in the same way, is consistent if it has at least three characters and $p$ and one of the identity forms are prefixes of each other. For the executable \texttt{lxsession-db} of the Debian project lxsession, for example, the candidate lxsession is consistent with the identity form \texttt{lxsessiondb}. All other name-only records, such as names that occur only in copyright notices, URLs, or symbol prefixes, form the name layer $O_f^{\mathrm{name}}$, which is reported apart from the main layer.

\paragraph{Self record.}
If $s(f) \neq \bot$ and no self record names the stem of $s(f)$, that is, $s(f)$ without its \texttt{lib} prefix and \texttt{.so} suffix, with or without the trailing numeric parts of the stem, the step adds a self record named after $s(f)$ without its \texttt{.so} suffix, such as \texttt{libgtk-x11-2.0} for \texttt{libgtk-x11-2.0.so.0}, with $v = \bot$. Every shared library thus reports its own identity, even when neither pass proposes it.

\paragraph{Versions.}
Three rules are applied, in this order, to the version of every reported record. First, a version whose every occurrence is a symbol-version name, such as \texttt{GLIBC\_2.34}, or the numeric suffix of a shared-library name, such as \texttt{libfoo.so.1.2}, is replaced by $\bot$, because these numbers are interface versions rather than releases. Second, a version that carries a distribution packaging marker, such as \texttt{+dfsg} or \texttt{+deb12u1}, loses its packaging suffix together with the Debian revision, so that \texttt{1.2.3+dfsg-2} becomes \texttt{1.2.3}. Third, a version that occurs in no string is replaced by $\bot$. Name-only records therefore always have $v = \bot$, and so does a dependency such as glibc whose only version evidence is its symbol-version requirements; the requirements themselves remain in $E$. For \texttt{wcsinfo}, the version 2.7 that the model derives from \texttt{libgsl.so.27} is cleared, because 2.7 occurs in the strings only in the symbol version \texttt{GLIBC\_2.7}. Records with the same product and version are finally merged across the two passes.

The main layer $O_f^{\mathrm{main}}$ thus consists of the self and embedded records, the dynamic-link layer $O_f^{\mathrm{dyn}}$ of the dynamic-link records, and the name layer $O_f^{\mathrm{name}}$ of the remaining name-only records; candidates in the two tiers marked as not reported are dropped. Every record keeps its tier $\tau$ and its evidence $E$, so that a consumer can apply a stricter acceptance, for example only same-string anchors, without rerunning the analysis.

\begin{algorithm}[t]
\caption{\TOOLNAME{} on a batch $F$ of ELF files}
\label{alg:relicbom}
\begin{algorithmic}[1]
\Require batch $F$; catalogue $P$ with version hints $V$
\Ensure records $O_f = O_f^{\mathrm{main}} \cup O_f^{\mathrm{dyn}} \cup O_f^{\mathrm{name}}$ for every $f \in F$
\For{each $f \in F$} \Comment{proposal}
    \State prepare $D(f)$, $S(f)$, and the evidence store; run Algorithm~\ref{alg:pass} in both modes
\EndFor
\State $G \gets$ generic names of the batch \Comment{Table~\ref{tab:ladder}}
\For{each $f \in F$} \Comment{acceptance}
    \For{each candidate $(p, v) \in R_{\mathrm{guided}}(f) \cup R_{\mathrm{open}}(f)$}
        \State $(\tau, E) \gets \Call{Tier}{p, v, f, G}$; \textbf{if} $\tau$ is not reported \textbf{then} skip $(p, v)$
        \State layer $\gets$ dynamic-link if $\tau$ is dynamic link; name if $\tau$ is name-only and $p$ is not consistent with $f$; else main
        \State add $(p, \Call{CleanVersion}{v}, \rho(\tau), \tau, E)$ to that layer, merging equal $(p, v)$
    \EndFor
    \State add the self record of $s(f)$ if $s(f) \neq \bot$ and no self record names its stem
    \State assign $\kappa$; look up vulnerabilities where $\kappa \neq \bot$ and $v \neq \bot$
\EndFor
\end{algorithmic}
\end{algorithm}

\paragraph{Catalogue annotation.}
Finally, each reported record receives the catalogue product $\kappa$: the product of $P$ whose name, reduced to lower-case letters and digits and taken with or without a leading \texttt{lib}, equals that of $p$, or $\bot$ if there is none. A record with $\kappa \neq \bot$ and $v \neq \bot$ is passed to the CVE scanner of CVE Binary Tool, which compares $v$ with the affected version ranges recorded for $\kappa$. The other records receive no vulnerability list but remain in the SBOM, so that a consumer sees which components lack vulnerability data instead of assuming that they have none. Algorithm~\ref{alg:relicbom} summarizes the complete procedure for a batch of files, from the two proposal passes of Algorithm~\ref{alg:pass} to the annotated records.

\section{Evaluation}
\label{sec:evaluation}

We evaluate \TOOLNAME{}{} through the following research questions:

\begin{itemize}
    \item \textbf{RQ1 (Effectiveness):} How effectively does \TOOLNAME{} recover components that appear in no reference?
    \item \textbf{RQ2 (Ablation):} How much does each design component contribute to \TOOLNAME{}?
    \item \textbf{RQ3 (Cost):} How much does it cost to run \TOOLNAME{}?
    \item \textbf{RQ4 (Practical use):} What does \TOOLNAME{} recover from real-world firmware?
\end{itemize}

\subsection{Experimental Setup}
\label{subsec:setup}

\paragraph{LLM setup.}
Every model call of \TOOLNAME{} uses \texttt{gpt-\allowbreak oss-\allowbreak 20b}, an open-weight model that can also run on local hardware, as the analysis of confidential firmware requires; in our experiments it is served through the Groq API at temperature~0. The evidence store uses the \texttt{nomic-embed-\allowbreak text} embedding model, served locally. The model sees each file only under its SHA-256 name, so that no stage can read the label off the file name; the acceptance step, which calls no model, uses the installed file name for the identity-consistency test (Section~\ref{subsec:ladder}), as it would on an unpacked image, and RQ2 quantifies its contribution.

\paragraph{Baselines.}
As baselines, we use four existing tools that cover the kinds of reference on which binary component identification relies. CVE Binary Tool~\cite{cvebintool} pairs a list of product names with hand-written signature checkers; its product names form the catalogue $P$ of \TOOLNAME{}. EMBA~\cite{emba} is an open-source firmware analyzer that identifies the components of binaries by version-identification rules, and represents the tools that auditors apply to firmware images. Syft~\cite{syft} is a widely used SBOM generator; besides package metadata, it recognizes a set of well-known binaries by built-in classifiers, and represents the generators that a practitioner would run first. BLADE~\cite{liu2026blade} serves a different purpose: it is a third-party library (TPL) detector, which matches the strings and function names of a binary against a feature database of 3,905 libraries to decide which known libraries the binary reuses, rather than a tool that recovers the SBOM of a binary. We include it because it is the most recent binary component detector and the closest to \TOOLNAME{}: like \TOOLNAME{}, it uses a language model, which selects among the retrieved candidates. Every tool runs in its default configuration and receives each file under its SHA-256 name, without package metadata and without network updates; BLADE uses the same language model as \TOOLNAME{}.

\paragraph{Dataset.}
RQ1 and RQ2 use a sample of Debian ELF files whose components appear in no reference. We drew it from a Debian archive mirror of 1,408,823 binary packages built from 40,887 source packages, restricted to four architectures (\texttt{amd64}, \texttt{arm64}, \texttt{armhf}, and \texttt{mipsel}). We selected the source packages by checking them against the reference of \TOOLNAME{}, the catalogue $P$, which consists of the 130,087 product names of CVE Binary Tool: a package entered the sampling frame only if neither its name nor the name of any of its binary packages matches a product in $P$ under the name normalization used for scoring. So that no evaluated tool holds the component in its reference, the same check was applied to the libraries of BLADE's feature database. Packages that occur in our development data were excluded as well. Because the components of firmware are mostly native C and C++ code, we also excluded the packages of language ecosystems, such as Python modules, Haskell libraries, and Go programs, following the naming conventions of Debian's language teams. This left 9,709 source packages, which a fixed seed puts in random order; the sample consists of the eligible projects among the first 1,000.

For each project we took its latest version on one of the four architectures and unpacked its binary packages without executing their contents. We kept executables and shared objects, and excluded interpreter extension modules and files built by the Go, Rust, GHC, or OCaml toolchains. Files were deduplicated by SHA-256 across the sample, and at most three files per project were chosen by a seeded hash rather than by name, so that the sample does not favor files that name their project. After unpacking, each project was checked again against the same references, this time also by the upstream name and the homepage recorded in its Debian metadata, and projects that matched were set aside. A final check found seven of the remaining 725 projects in a reference under another name, such as \texttt{snowball} as \texttt{libstemmer}; we excluded them before \TOOLNAME{} or any other LLM-based tool had been run on the sample. The test set thus consists of 1,173 files from 718 projects. \TOOLNAME{} processes the files of all eligible projects, 1,517 in total, as one batch (Algorithm~\ref{alg:relicbom}).

The label of a file is its Debian source package, with the names of its binary packages and the upstream name from its copyright file as aliases. Its version label is the Debian version without the epoch, the Debian revision, and repackaging suffixes such as \texttt{+dfsg}. Snapshot versions, marked by version-control tokens or eight-digit dates, receive no version label, which leaves 1,066 version-labelled files. The labels are derived automatically from package metadata; they identify one known component per file and are not complete SBOMs. Inputs and labels were frozen before any tool was run on the sample.

\paragraph{Firmware.}
RQ4 uses 101 proprietary firmware images from network security appliances and two automotive vendors. They differ in packaging format, file-system layout, and run-time environment, and include custom real-time operating system layers, vendor-specific modules, and statically linked executables. No component labels exist for these images. Because they are proprietary, they are not part of the released artifact.

\subsection{RQ1: Effectiveness}
\label{subsec:rq1}

\paragraph{What is measured.}
Each file of the test set belongs to one Debian project, which we call the \emph{target} of the file. In a deployed image, most components appear as files of their own, such as a server executable or a shared library, so naming the project that built each file is the core of the image's SBOM, and it is exactly what file names and the missing package metadata no longer reveal. Debian records the target of every file, which makes it a reliable label; the tools see only the binary. For example, the executable \texttt{wcsinfo} belongs to Astrometry.net 0.93. Its strings state that ``this program is part of the Astrometry.net suite'' and contain the release \texttt{0.93}, and \TOOLNAME{} reports \texttt{astrometry.net} 0.93 in the main layer, together with \texttt{qfits} 6.2.0, a FITS library that Astrometry.net bundles in its source tree. It reports the libraries that the file loads, such as GSL and the C library, as dynamic-link records, and \texttt{anqfits}, which the binary only names, in the name layer.

As is common for binary component detection~\cite{liu2026blade}, we measure recall over (file, component) pairs, with the counts aggregated over all files. Debian records the target of each file but not the other projects that the file embeds, so recall is measured on the target. \emph{Component recall} is the share of all files for which a main-layer record names the target. \emph{Version accuracy} is measured on the targets that are recovered: among the version-labelled files whose target a main-layer record names, it is the share for which that record carries the labelled release, and a record without a release counts as incorrect. We measure the release separately because vulnerabilities are recorded against releases: a component without its release, or with a wrong one, cannot be checked for known vulnerabilities. Names are compared after case folding and with the aliases of the label, and releases as strings, so that 3.1 does not match 3.10. The dynamic-link and name layers do not count, because their records do not state that the file contains the component. The other main-layer records, such as \texttt{qfits} above, cannot be scored against the labels; we assess them by \emph{precision}, the share of the reported main-layer records that are correct, and report the number of main-layer records per file as the amount of output that a user has to review.

\begin{table*}[t]
\centering
\caption{Results on the test set, overall and by architecture (number of files in parentheses). R: component recall over all files; VA: version accuracy over the version-labelled files whose target the tool recovers; P: precision of the reported records, assessed for \TOOLNAME{} only. The version accuracy of EMBA and BLADE rests on one and three recovered targets.}
\label{tab:rq1}
\small
\setlength{\tabcolsep}{4pt}
\begin{tabular}{@{}lccc@{\hspace{6pt}}cc@{\hspace{6pt}}cc@{\hspace{6pt}}cc@{\hspace{6pt}}cc@{}}
\toprule
& \multicolumn{3}{c}{Overall (1,173)} & \multicolumn{2}{c}{\texttt{amd64} (397)} & \multicolumn{2}{c}{\texttt{arm64} (261)} & \multicolumn{2}{c}{\texttt{armhf} (265)} & \multicolumn{2}{c}{\texttt{mipsel} (250)} \\
\cmidrule(lr){2-4}\cmidrule(lr){5-6}\cmidrule(lr){7-8}\cmidrule(lr){9-10}\cmidrule(l){11-12}
Tool & R & VA & P & R & VA & R & VA & R & VA & R & VA \\
\midrule
CVE Binary Tool & 0.0 & -- & -- & 0.0 & -- & 0.0 & -- & 0.0 & -- & 0.0 & -- \\
EMBA & 0.1 & 100.0 & -- & 0.0 & -- & 0.4 & 100.0 & 0.0 & -- & 0.0 & -- \\
Syft & 0.0 & -- & -- & 0.0 & -- & 0.0 & -- & 0.0 & -- & 0.0 & -- \\
BLADE & 0.3 & 33.3 & -- & 0.3 & 0.0 & 0.8 & 50.0 & 0.0 & -- & 0.0 & -- \\
\midrule
\rowcolor{gray!12}
\TOOLNAME{} & \textbf{39.6} & \textbf{44.5} & \textbf{78.7} & \textbf{39.8} & \textbf{46.2} & \textbf{43.7} & \textbf{39.8} & \textbf{34.3} & \textbf{56.2} & \textbf{40.4} & \textbf{36.7} \\
\bottomrule
\end{tabular}
\end{table*}

\paragraph{Results.}
Table~\ref{tab:rq1} reports the results. \TOOLNAME{} recovers the target of 464 of the 1,173 files (39.6\%), with 1.10 main-layer records per file. Because the files of one project are not independent, we resample projects to obtain a confidence interval (2,000 bootstrap resamples), which is 36.2--43.0\%. Of the recovered targets, 418 have a version label, and for 186 of them \TOOLNAME{} reports the labelled release (version accuracy 44.5\%). When it reports a release at all, the release is correct in 78.2\% of the cases (186 of 238). Because the test set was drawn so that its components appear in no reference, the baselines are not expected to recover them, and they do not: CVE Binary Tool and Syft recover none of the components, EMBA one, and BLADE three.

\paragraph{Precision.}
Precision is usually computed against complete dependency lists, which count every record outside the list as wrong~\cite{liu2026blade}. Our labels name only the target, so a record that does not match it, such as \texttt{qfits} for \texttt{wcsinfo}, is not necessarily wrong. We therefore judge the main-layer records of \TOOLNAME{} against the binary. Of its 1,289 records, four merely restate a dynamic dependency and are left out, and 128 name a component of their file a second time, such as \texttt{libo2.so} next to \texttt{libo2}, and are merged with the first, so that no component is counted twice. Of the remaining 1,157 records, 771 name the target or the file itself, as identified by its \texttt{DT\_SONAME} or file name, and count as correct; five name an operating system or a name of at most two characters and count as incorrect. For the other 381 records, the binary shows the name but not whether the component is embedded or only mentioned. One author judged each of them against the strings of the file, its dynamic dependencies, the copyright file of its package, and the package source, following written criteria that we release with the artifact. In 140 of these records, the code of the named component is in the file; they include the project under another name, parts of the project, and bundled third-party libraries. Counting the other 241 as incorrect, the precision of \TOOLNAME{} is 78.7\% (911 of 1,157). We do not estimate the precision of the baselines: they recover almost none of the labelled components, and the string-based checks cannot judge records that rest on signatures or function names.

\paragraph{Evidence behind the recovered components.}
Every component that \TOOLNAME{} recovers is accepted on evidence in the file (Section~\ref{subsec:ladder}). Table~\ref{tab:rq1-evidence} divides the 464 recovered components by the strongest evidence among the records that name them. For 98 components, the file declares the component as its own identity in \texttt{DT\_SONAME}. For 149, a single string names the component together with a release, as a version banner or a versioned build path does. For 95, the name and the release appear in different strings and are bound by the cross-string rule, and for 122, the binary names the component without a release and the name matches the file's own name. This evidence is often not visible in the file name: \texttt{libplmsys.so.1.9}, for instance, is identified as Plastimatch 1.9.4 from its banner, although its name suggests neither the project nor the release.

\begin{table}[t]
\centering
\caption{Evidence behind the 464 components that \TOOLNAME{} recovers on the test set, with an example for each kind.}
\label{tab:rq1-evidence}
\small
\begin{tabular}{@{}>{\raggedright\arraybackslash}p{0.17\linewidth}r>{\raggedright\arraybackslash}p{0.17\linewidth}>{\raggedright\arraybackslash}p{0.16\linewidth}>{\raggedright\arraybackslash}p{0.28\linewidth}@{}}
\toprule
Evidence & Comp. & Example file & Record & Evidence in the file \\
\midrule
Self (\texttt{DT\_SONAME}) & 98 & \texttt{libquvi-0.9-\allowbreak 0.9.4.so} & libquvi 0.9.4 & \texttt{DT\_SONAME} \texttt{libquvi-0.9-\allowbreak 0.9.4.so} \\
\addlinespace
Name and release in one string & 149 & \texttt{bgpq4} & bgpq4 1.9 & ``bgpq4 version: 1.9 (https://\allowbreak github.com/bgp/bgpq4)'' \\
& & \texttt{libplmsys.\allowbreak so.1.9} & plastimatch 1.9.4 & ``Plastimatch 1.9.4'' \\
\addlinespace
Name and release in different strings & 95 & \texttt{wcsinfo} & astrometry.net 0.93 & ``This program is part of the Astrometry.net suite.'' and ``0.93'' \\
\addlinespace
Name only, matching the file's name & 122 & \texttt{lxsession-db} & lxsession (no release) & ``- Lxsession database utility'' 
\\
\bottomrule
\end{tabular}
\end{table}

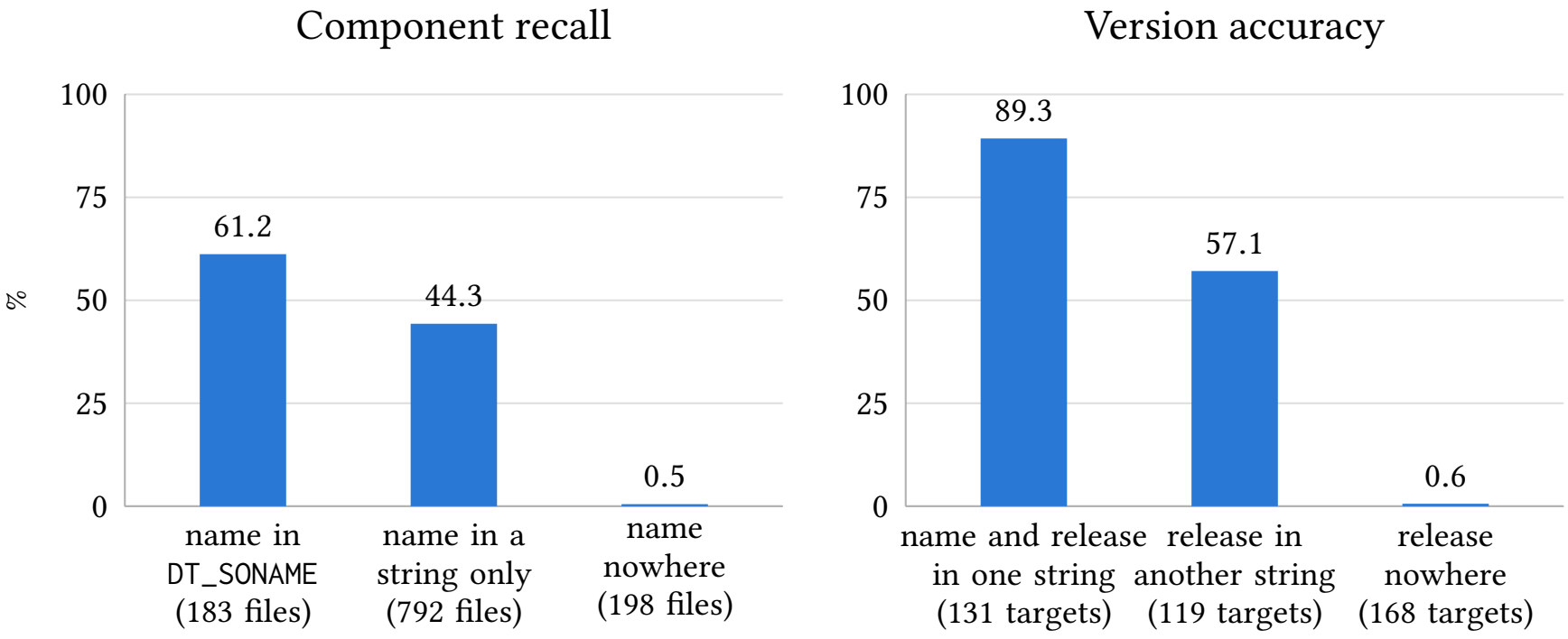
\begin{figure}[t]
\centering
\definecolor{relicblue}{HTML}{2A78D6}
\pgfplotsset{relicbars/.style={
  ybar, bar width=18pt, width=6.4cm, height=4.6cm, ymin=0, ymax=100, ytick={0,25,50,75,100},
  xtick={1,2,3}, enlarge x limits=0.28, axis x line*=bottom, axis y line*=left,
  ymajorgrids, grid style={gray!25}, axis line style={gray!60}, major tick length=0pt,
  tick label style={font=\scriptsize}, label style={font=\scriptsize}, title style={font=\small},
  x tick label style={align=center, text width=2.1cm},
  nodes near coords, nodes near coords style={font=\scriptsize, text=black},
  every node near coord/.append style={/pgf/number format/.cd, fixed, fixed zerofill, precision=1}}}
\begin{tikzpicture}
\begin{axis}[relicbars, name=comp, ylabel={\%}, title={Component recall},
  xticklabels={{name in\\\texttt{DT\_SONAME}\\(183 files)}, {name in a\\string only\\(792 files)}, {name\\nowhere\\(198 files)}}]
\addplot[fill=relicblue, draw=none] coordinates {(1,61.2) (2,44.3) (3,0.5)};
\end{axis}
\begin{axis}[relicbars, at={(comp.east)}, anchor=west, xshift=0.9cm, title={Version accuracy},
  xticklabels={{name and release\\in one string\\(131 targets)}, {release in\\another string\\(119 targets)}, {release\\nowhere\\(168 targets)}}]
\addplot[fill=relicblue, draw=none] coordinates {(1,89.3) (2,57.1) (3,0.6)};
\end{axis}
\end{tikzpicture}
\caption{Results of \TOOLNAME{} on the test set by the evidence that each file contains. Left: component recall over all 1,173 files. Right: version accuracy over the 418 recovered targets with a version label.}
\label{fig:rq1-strata}
\end{figure}

\paragraph{Recall and the evidence in the binary.}
Recovery from evidence can succeed only where the binary contains that evidence, and Figure~\ref{fig:rq1-strata} shows the results by the evidence that each file carries. Where the file names its component in \texttt{DT\_SONAME}, \TOOLNAME{} recovers 61.2\% of the components, and where only a printable string names it, 44.3\%. Taken together, it recovers 463 of the 975 components that their binaries name (47.5\%); the remaining 198 files never name their project and lie outside the scope of \TOOLNAME{} (Section~\ref{subsec:formulation}). Releases follow the same pattern. Among the 418 recovered targets with a version label, \TOOLNAME{} reports the labelled release for 89.3\% when one string carries both the name and the release, and for 57.1\% when the release appears apart from the name. For 168 recovered targets, the release appears in no string of the binary, and for 132 of them \TOOLNAME{} reports no release rather than guessing one, as the design intends (Section~\ref{subsec:formulation}).

\paragraph{Robustness across architectures and file types.}
Because \TOOLNAME{} reads the text and the dynamic segment of a binary rather than its machine code, its recall is not expected to depend on the instruction set. Across the four architectures of the test set (Table~\ref{tab:rq1}), component recall ranges from 34.3\% (\texttt{armhf}) to 43.7\% (\texttt{arm64}) and version accuracy from 36.7\% (\texttt{mipsel}) to 56.2\% (\texttt{armhf}), including the two 32-bit architectures that are common in embedded devices. Recall is also similar for shared objects with a \texttt{DT\_SONAME} (41.0\%) and for files without one, such as executables and plugins (39.0\%), so the result does not rest on the self record of shared objects.

\smallskip
\noindent\fbox{\parbox{\dimexpr\linewidth-2\fboxsep-2\fboxrule}{\textbf{Answer to RQ1:} On binaries whose components appear in no reference, \TOOLNAME{} recovers the component of 39.6\% of the files, whereas the baselines recover at most 0.3\%, reports the correct release for 44.5\% of the components it recovers, and 78.7\% of its records are correct. Every recovered component rests on evidence in the file: where the binary names its component, \TOOLNAME{} recovers 47.5\% of the components, and where one string carries the name and the release, it reports the correct release for 89.3\% of them. The results are stable across four architectures.}}

\subsection{RQ2: Ablation}
\label{subsec:rq2}

We ablate the three parts of \TOOLNAME{} in the order of Section~\ref{sec:methodology}. For candidate proposal (Section~\ref{subsec:proposal}) and evidence-gated acceptance (Section~\ref{subsec:ladder}), we keep the proposals of both passes fixed and repeat only what follows them. The acceptance step is deterministic and calls no model, so each variant reuses the stored model outputs on the test set and differs from the full design in exactly one component; removing a pass means accepting the candidates of the other pass alone. Evidence preparation (Section~\ref{subsec:evidence}) changes what the model sees and therefore needs model runs. We compare two runs of the catalogue-guided pipeline with the same model that differ only in how the evidence reaches each stage: one passes a ranked selection of at most 40 strings of 150 characters, as the first version of \TOOLNAME{} did, and the other uses the evidence store and the per-stage retrieval of Section~\ref{subsec:evidence}. Neither run includes the open pass or the acceptance step, and the first was stopped after 1,149 of the 1,517 files, so we compare them on the 888 test files that both completed. We report component recall and version accuracy as in RQ1, the number of records per file, and two kinds of record that the automatic checks of RQ1 identify without manual review: records that are provably wrong, because they name an operating system, a compiler, a name of at most two characters, or a name that occurs in no string of the binary, and records that merely restate a dynamic dependency. Table~\ref{tab:rq2} shows the results.

\begin{table}[t]
\centering
\caption{Ablation by the parts of Section~\ref{sec:methodology}. R: component recall (difference to the reference of the group in parentheses); VA: version accuracy; Rec./file: records per file; Wrong: records that the automatic checks prove wrong; Dep.: dynamic dependencies reported as components. The evidence-preparation group compares two runs of the guided pipeline without the acceptance step on 888 files; the other groups re-run the acceptance step of the full design on all 1,173 files.}
\label{tab:rq2}
\small
\setlength{\tabcolsep}{4pt}
\begin{tabular}{@{}lrrrrr@{}}
\toprule
Variant & R (\%) & VA (\%) & Rec./file & Wrong & Dep. \\
\midrule
\multicolumn{6}{@{}l}{\textit{Evidence preparation (Section~\ref{subsec:evidence}); guided pipeline, 888 files}} \\
\quad with retrieval (reference) & 10.0 & 37.5 & 2.91 & 49 & 1,926 \\
\quad without retrieval & 5.4 (\,$-$4.6) & 81.4 & 0.90 & 28 & 563 \\
\midrule
\rowcolor{gray!12}
\TOOLNAME{}, full design (reference) & \textbf{39.6} & 44.5 & 1.10 & 5 & 4 \\
\addlinespace
\multicolumn{6}{@{}l}{\textit{Candidate proposal (Section~\ref{subsec:proposal})}} \\
\quad without the open pass & 9.5 (\,$-$30.1) & 31.2 & 0.51 & 1 & 2 \\
\quad without the guided pass & 38.4 (\,$-$1.2) & 44.8 & 0.96 & 4 & 2 \\
\addlinespace
\multicolumn{6}{@{}l}{\textit{Evidence-gated acceptance (Section~\ref{subsec:ladder})}} \\
\quad without the evidence ladder & 54.7 (+15.1) & 34.9 & 6.47 & 93 & 5,016 \\
\quad without identity consistency & 29.2 (\,$-$10.4) & 60.0 & 0.84 & 5 & 3 \\
\quad without the self record & 38.8 (\,$-$0.8) & 45.5 & 1.00 & 5 & 4 \\
\quad without the version rules & 39.6 (\,\phantom{$-$}0.0) & 44.0 & 1.13 & 5 & 4 \\
\bottomrule
\end{tabular}
\end{table}

\paragraph{Evidence preparation.}
Retrieval nearly doubles the targets that the guided pipeline recovers, from 48 to 89 of the 888 files (5.4\% to 10.0\%), because each stage sees the blocks chosen for its own candidates instead of one ranked selection that has to serve every stage. The wider evidence also brings more records: 2.91 instead of 0.90 per file, most of them dynamic dependencies, and more wrong releases, so that version accuracy falls from 81.4\% to 37.5\% and the correct releases from 35 to 30. In the full design, the acceptance step sorts these records: dependencies go to the dynamic-link layer, and a release that the binary does not support is discarded (Section~\ref{subsec:ladder}).

\paragraph{Candidate proposal.}
The open pass provides most of the recall. Without it, \TOOLNAME{} recovers 9.5\% of the targets, 353 fewer, because the guided pass maps the proposed names onto the catalogue, which by construction of the test set does not contain the targets; it then behaves much like a tool bounded by its catalogue. The guided pass adds 14 targets on the test set. Its role lies with components that the catalogue lists, for which it supplies the catalogue spelling and the version hints (Section~\ref{subsec:proposal}), and the test set excludes such components by construction.

\paragraph{Evidence-gated acceptance.}
Without the evidence ladder, every proposed record enters the main layer. Component recall rises to 54.7\%, but the output grows from 1.10 to 6.47 records per file, 5,016 of them dynamic dependencies presented as embedded components and 93 provably wrong, and version accuracy falls from 44.5\% to 34.9\%. The targets that only this variant places in the main layer are not lost in the full design: of these 187 targets, 79 are reported in the dynamic-link layer, because the file loads a library of its own project rather than containing its code, and 108 in the name layer, because the binary names them without evidence that their code is in the file. The ladder thus decides where each candidate belongs, which keeps the main layer to the components that the file contains. Identity consistency admits 122 targets (10.4 points) that the binary names without a release, when the name matches the file's own name; the model never sees the file name, and the acceptance step uses it only to admit names that the model proposed from the content of the file, while \texttt{DT\_SONAME} alone would admit 347 targets (29.6\%) in total. Version accuracy rises without identity consistency only because the removed records carry no release. The self record from \texttt{DT\_SONAME} adds nine targets, and the version rules, which discard releases that the binary shows only as a symbol version or a shared-object number, or not at all, and strip Debian packaging suffixes, reduce the wrong releases among the recovered targets from 61 to 52 and turn two of them into correct ones.

\smallskip
\noindent\fbox{\parbox{\dimexpr\linewidth-2\fboxsep-2\fboxrule}{\textbf{Answer to RQ2:} Each part of \TOOLNAME{} contributes. Retrieval nearly doubles what the guided pipeline recovers. In candidate proposal, the open pass is the main source of recall beyond references (30.1 points). In acceptance, identity consistency adds another 10.4 points, and the evidence ladder keeps the output to 1.10 records per file instead of 6.47 and keeps 5,016 dynamic dependencies and 93 provably wrong records out of the main layer, while the targets it does not accept as embedded remain in the dynamic-link and name layers.}}

\subsection{RQ3: Cost}
\label{subsec:rq3}

We take the cost from the two runs that produce the results of RQ1. For every file, they record the model calls, the tokens that the provider reports, the time spent building the evidence store, and the time spent waiting for the model; the acceptance step calls no model. Both runs used \texttt{gpt-oss-20b} served through the Groq API with four parallel workers. Because firmware sometimes cannot be shared with a model provider, we also report a run of the guided pipeline with retrieval on a single local GPU, with the same model served by Ollama, on 478 files of our development data. From these measurements, we estimate the time and cost of analyzing whole images.

\begin{table}[t]
\centering
\caption{Cost of \TOOLNAME{} per file on the test set (1,173 files; Groq API, four parallel workers). Prices at the list price of October 2026.}
\label{tab:rq3}
\small
\begin{tabular}{@{}lrrr@{}}
\toprule
& Guided pass & Open pass & Both passes \\
\midrule
Model calls & 2.92 & 2.93 & 5.85 \\
Prompt tokens & 10,891 & 10,705 & 21,596 \\
Completion tokens & 5,372 & 5,051 & 10,423 \\
\quad of which reasoning & 5,132 & 4,752 & 9,884 \\
Time per file, mean (s) & 15.0 & 17.5 & 32.5 \\
\quad median / 90th percentile (s) & 10.3 / 25.1 & 10.8 / 37.4 & 23.3 / 68.5 \\
\quad building the evidence store (s) & 1.6 & 1.7 & 3.3 \\
\quad waiting for the model (s) & 8.8 & 12.0 & 20.8 \\
Cost per file (US\$) & 0.0024 & 0.0023 & 0.0047 \\
\addlinespace
Cost per recovered component (US\$) & & & 0.012 \\
Cost per correct release (US\$) & & & 0.030 \\
\bottomrule
\end{tabular}
\end{table}

Table~\ref{tab:rq3} shows the cost of a file. Across its two passes, \TOOLNAME{} issues 5.9 model calls per file. A file consumes about 21,600 prompt tokens and 10,400 completion tokens. Of the completion tokens, 95\% are the hidden reasoning of the model, so the cost depends more on how long the model reasons than on how much evidence it reads. At the list price of the provider, a file costs less than half a cent (US\$0.0047), and the 1,173 test files cost US\$5.57 in total. Measured by what it delivers, \TOOLNAME{} spends about 1.2 cents for each component that it recovers and about 3 cents for each correct release.

The calls follow the stages of Section~\ref{subsec:proposal}. Every file receives a proposal in both passes. In the guided pass, 87\% of the files go on to review and 75\% to synthesis, because a proposed name must also match the catalogue to reach them; in the open pass, about 94\% reach both. The proposal is the most expensive stage: the model writes its longest answers there, while the later stages, which see only the evidence retrieved for a few candidates, produce shorter ones. Building the evidence store takes less than two seconds per file and calls no model.

Binaries differ widely in size: the test set ranges from files with a few dozen printable strings to a file with 570,000. The prompt nonetheless stays bounded. If we divide the files into quarters by the number of their strings, the prompt grows from about 8,200 tokens per file in the smallest quarter to about 30,000 in the third, and then stops growing: the largest quarter, whose files have a median of 5,350 strings, uses no more prompt tokens than the third quarter, whose median is 928. Each stage receives at most a fixed number of blocks and characters (Section~\ref{subsec:evidence}), so the prompt is bounded by the retrieval budgets rather than by the size of the file, and no single large binary can exhaust the context of the model. Over both passes, the cheapest file needs about 1,300 prompt tokens and the most expensive about 65,000. The time per file still grows, from 17.8 seconds in the smallest quarter to 51.6 seconds in the largest, mostly because building the evidence store takes longer for large binaries.

Most of the time of a file is spent waiting for the model rather than computing, so files can be analyzed in parallel. With four workers, the guided pass processed the 1,517 files of the sample in 97 minutes, or 3.8 seconds per file, which matches its mean time per file divided by the number of workers (3.75 seconds). By the same relation, both passes with a hosted model and four workers analyze an image of 100 ELF files in about 14 minutes and an image of 1,000 ELF files in about 2.3 hours, for US\$0.47 and US\$4.7 respectively; more workers shorten the time further, within the rate limits of the provider.

\TOOLNAME{} needs no reference to be built before it analyzes an image. Reference-based detectors first compile or summarize the libraries that they can recognize into a feature database, and they must extend it as new libraries and releases appear (Section~\ref{sec:background}). The only prepared input of \TOOLNAME{} is its catalogue, the product list of a vulnerability scanner, which a vulnerability assessment needs anyway; the evidence store is built for each file as part of its analysis. The first analysis of an image from an unfamiliar vendor therefore costs the same as any other, and a component that no reference has seen costs nothing extra to look for.

\smallskip
\noindent\fbox{\parbox{\dimexpr\linewidth-2\fboxsep-2\fboxrule}{\textbf{Answer to RQ3:} With a hosted model, \TOOLNAME{} costs less than half a cent and about half a minute per file, or about 1.2 cents for each component that it recovers. Its prompts stay bounded however large the binary, because retrieval caps the evidence that each stage receives, and parallel analysis covers an image of 1,000 ELF files in about two hours. It needs no reference to be built in advance, and on a single local GPU it analyzes firmware that must not leave the machine at about a minute per file.}}

\subsection{RQ4: Practical Use}
\label{subsec:rq4}

To answer RQ4, we apply \TOOLNAME{}, CVE Binary Tool, and EMBA to the ELF files of the 101 firmware images. We use these two baselines because they are intended for the analysis of firmware binaries, and we compare the tools at the level of normalized products, because they report at different granularities. No component labels exist for the images, so we measure what \TOOLNAME{} adds to the baselines: the products that only \TOOLNAME{} reports, and whether the binaries support them.

\begin{table}[t]
\centering
\caption{Products reported by \TOOLNAME{} on the firmware images and their overlap with the baselines.}
\label{tab:SBOM_coverage}
\small
\begin{tabular}{@{}lr@{}}
\toprule
Measure & Value \\
\midrule
Records reported by \TOOLNAME{} & 7,576 \\
Distinct products in these records & 193 \\
\quad also reported by CVE Binary Tool & 37 (19\%) \\
\quad also reported by EMBA & 28 (14\%) \\
\quad also reported by both baselines & 20 (10\%) \\
\quad \textbf{reported by neither baseline} & \textbf{148 (77\%)} \\
\bottomrule
\end{tabular}
\end{table}

\begin{figure}[t]
    \centering
    \includegraphics[width=0.8\linewidth]{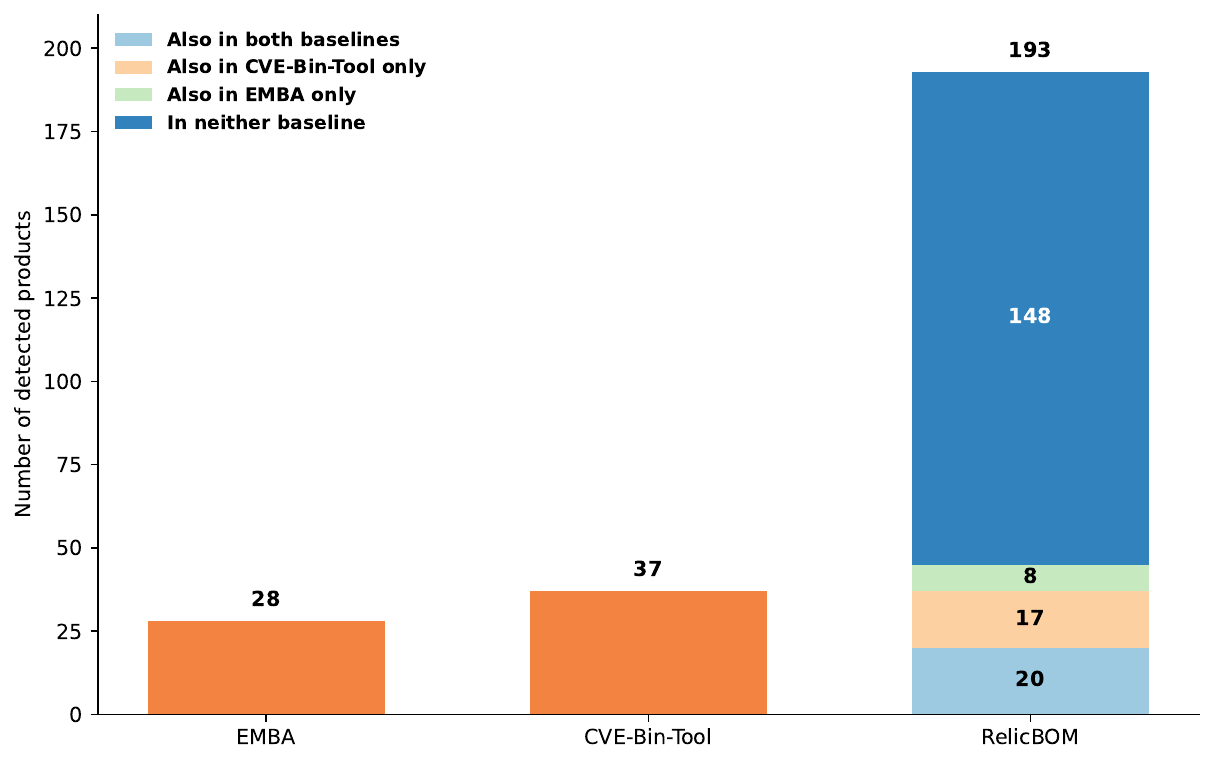}
    \caption{Products reported on the firmware images. EMBA and CVE Binary Tool are shown as the totals of the products they report; the products of \TOOLNAME{} are divided by whether a baseline also reports them.}
    \label{fig:comparison}
\end{figure}

Table~\ref{tab:SBOM_coverage} summarizes the outcome. \TOOLNAME{} produced 7,576 records, which name 193 distinct products after normalization. CVE Binary Tool also reports 37 of these products, EMBA 28, and both baselines 20. This shared part consists of ubiquitous system components such as \texttt{bzip2}, \texttt{zlib}, and \texttt{openssl}, indicating that \TOOLNAME{} also reports the common building blocks that the baselines identify. The remaining 148 products, 77\% of those that \TOOLNAME{} reports, appear in neither baseline's output (Figure~\ref{fig:comparison}). The records that name them rest on evidence in the binaries, such as banners, library identifiers, and version-bearing strings; we characterize these products and validate them below.

\paragraph{Characteristics of the additional products.}
The additional products cover several layers of firmware composition. Vendor-specific components, such as \texttt{codesys:gateway}, expose software tied to particular platforms and industrial ecosystems. Connectivity libraries, including \texttt{bonjour}, \texttt{mdnsresponder}, and \texttt{google-protobuf}, reveal service discovery and communication dependencies. Smaller open-source libraries, such as \texttt{giflib} and \texttt{multimon-ng}, extend the inventory beyond the common system components shared with the baselines. Other products give a finer view of larger frameworks, such as \texttt{libavcodec} within FFmpeg and \texttt{alsa-lib} within ALSA, which identifies the particular module that a subsequent security analysis has to examine.

\paragraph{Validation of the additional products.}
Because no baseline reports these 148 products, two security researchers validated them manually instead of treating them as correct because \TOOLNAME{} reported them. For each product, they inspected the evidence retained from the binary, including file names, version banners, paths, library identifiers, log strings, and model identifiers, and checked that the recovered product and version follow from this evidence rather than from a superficial catalogue match. Under this protocol, all 148 products are supported by evidence in the binaries. Because the images include no source code, build metadata, or vendor SBOM, this result validates the evidence behind each record; it is not a comparison with a complete SBOM.

Table~\ref{tab:validation} shows two cases in which the file name alone says little about the component. \texttt{libIMGegl.so} suggests a graphics library, but its strings contain QNX-specific evidence and the release token \texttt{qnx6.5.0-sp1}, from which \TOOLNAME{} recovers \texttt{neutrino\_rtos} 6.5.0-sp1. In \texttt{Sirius\_SDARS\_Service}, the file name already indicates the product, but the binary contains several version-like strings; \TOOLNAME{} selects \texttt{2.00}, which occurs twice, for the main Sirius service and does not take internal component tokens such as \texttt{sxe\_dl\_v1.16.00}. Such records cannot be obtained by matching file names or raw strings alone.

\begin{table*}[t]
\centering
\caption{Representative manual validation examples for products that only \TOOLNAME{} reports.}
\label{tab:validation}
\small
\begin{tabular}{@{}>{\raggedright\arraybackslash}p{0.16\textwidth}>{\raggedright\arraybackslash}p{0.17\textwidth}>{\raggedright\arraybackslash}p{0.42\textwidth}>{\raggedright\arraybackslash}p{0.13\textwidth}@{}}
\toprule
Recovered record & File & Key evidence in the binary & Validation decision \\
\midrule
\texttt{neutrino\_rtos} 6.5.0-sp1 & \texttt{libIMGegl.so} & Strings include \texttt{ne-service-\allowbreak pasa-syn3-\allowbreak -qnx6.5.0-sp1} and \texttt{qnx software sys}. The token \texttt{qnx6.5.0-sp1} directly supports the version, and QNX Neutrino RTOS maps to \texttt{neutrino\_rtos}. & Product and version supported \\
\addlinespace
\texttt{sirius} 2.00 & \texttt{Sirius\_\allowbreak SDARS\_\allowbreak Service} & The file name indicates the Sirius product. The token \texttt{2.00} occurs twice in the strings; other version-like strings, such as \texttt{sxe\_dl\_v1.16.00}, belong to internal components rather than to the main Sirius service. & Product and version supported \\
\bottomrule
\end{tabular}
\end{table*}

\subsection{Discussion}
\label{sec:discussion}

Our results bear on those who use SBOMs recovered from binaries, on those who build analyses around language models, and on those who build the binaries; we close by considering executable formats other than ELF.

For those who use a recovered SBOM, each record is an inference, and they need to know how strongly the binary supports it. \TOOLNAME{} therefore reports, with every record, its relationship to the file, the evidence tier on which it was accepted, and the strings and loader entries behind it (Section~\ref{subsec:ladder}). Both SBOM standards provide for these fields: the relationship maps onto the SPDX types \texttt{CONTAINS} and \texttt{DYNAMIC\_LINK}, and the tier and its evidence onto the \texttt{identity} and \texttt{occurrences} entries of a CycloneDX \texttt{evidence} object (Section~\ref{subsec:bg-sbom}), so a recovered SBOM can be exchanged with existing tools. The evidence also lets users decide how far to rely on a record. The reported release is correct for 89.3\% of the recovered targets whose binary carries the name and the release in one string, and for 57.1\% of those whose release appears in a separate string (Section~\ref{subsec:rq1}). An assessment that must avoid false alarms can therefore match the former directly and confirm the latter, simply by filtering the tiers, without rerunning the analysis.

An undefined version is likewise information rather than a gap. For 132 of the 168 recovered targets whose release appears in no string, \TOOLNAME{} reports no release instead of guessing one. The assessment then knows that the component is present and that its release must be established in another way, for example from the vendor's release notes, instead of silently matching the component against the vulnerabilities of a release that is not installed. The other records serve other purposes. Within an image, the library named by a dynamic-link record is normally another file of the image with records of its own, so the records of all files form a dependency graph rather than a flat list. The name layer, which holds components that the binary only mentions, such as projects named in copyright notices, is not part of the inventory but a starting point for manual review.

For analyses built around language models, our results suggest a division of labor. A language model recognizes components in banners, paths, and log messages that no detection rule anticipates: the open pass, in which the model names components without a catalogue, contributes 30.1 points of recall (Section~\ref{subsec:rq2}). If every proposal were accepted, however, the output would grow to 6.47 records per file, most of them libraries that the file merely loads. The model should therefore propose, but not decide. Because acceptance in \TOOLNAME{} is deterministic and consults only the binary, a component whose name the binary does not contain is never reported, every record can be checked against the evidence it cites without trusting the model, and the acceptance step can be repeated, audited, or tightened on the stored proposals without new model calls, as our ablation did. The catalogue of a vulnerability scanner, in turn, no longer bounds what can be found, but helps to spell the names it lists and to attach vulnerabilities to the records it covers (Section~\ref{subsec:proposal}). We expect this division of labor to carry over to other tasks in which a model labels an artifact whose content can be checked mechanically.

For those who build binaries, our results show that recovery succeeds where the binary identifies itself. \TOOLNAME{} recovers 61.2\% of the components whose files name them in \texttt{DT\_SONAME}, whereas a file that names its project nowhere, as 198 test files do, can be recognized only by a reference that already contains its code. Whether a binary identifies itself is decided when it is built. A version banner with the project and its release, a library name derived from the project name, or a package note costs little to add and survives stripping, and the metadata conventions of Go, \texttt{cargo-auditable}, and the UAPI package notes record the identity of a file or of its embedded modules in machine-readable form (Section~\ref{subsec:bg-evidence}), which reduces identification to reading it. For producers who must supply SBOMs, embedding the identity at build time costs far less than recovering it afterwards. The binaries already deployed lack such metadata, and their components must be recovered from the evidence they retain, which is the setting that \TOOLNAME{} addresses.

Finally, both kinds of evidence that \TOOLNAME{} uses exist beyond ELF. A Windows PE file lists its imported libraries in its import directory and can declare its product name and version in a version resource; a Mach-O library states its install name and current version in a load command and lists the libraries it loads in further ones. These structures play the roles of \texttt{DT\_NEEDED} and \texttt{DT\_SONAME}, and a version resource, whose fields are defined to name the product and the release, can anchor a record more firmly than a string. Since residual text, the retrieval over it, and the model stages that read it do not depend on the format, supporting another format requires only a parser for its structures and a mapping of them onto the tiers of the evidence ladder. For real-time operating system firmware, which often consists of a single image without any loader structure, acceptance would rest on the string tiers alone.

\section{Limitations and Threats to Validity}
\label{sec:threats}

Our labels come from Debian package metadata and name one component per file, the project that built it (Section~\ref{subsec:setup}). They are reliable for this target but are not complete SBOMs, so recall measures only whether \TOOLNAME{} recovers the target, and the components that a file embeds enter the evaluation only through precision. For 381 of the records that do not match the target, the binary alone could not show whether the component is embedded, and one author judged them following written criteria that we release with the artifact. Such judgments involve interpretation, in particular for parts of a project and for projects that appear under other names. Names are compared after case folding and with the aliases of each label, so a correct record under a name that the label does not list counts as a miss, which makes the reported recall conservative. A component counts as outside every reference when no name, alias, homepage, or repository of its project matches the references of the evaluated tools; a reference may still contain it under an unrelated name, and we removed the seven projects that our final check found in this way. Finally, the baselines are built to recognize the components that their references contain, and the test set contains none of these by construction. Their low recall on it measures the gap that \TOOLNAME{} addresses.

Moreover, the test set consists of native executables and shared objects from Debian, which builds its packages from source with standard toolchains. It covers four architectures, including the two 32-bit architectures common in embedded devices, and the results are similar across them (Section~\ref{subsec:rq1}). Firmware binaries are often older, built with vendor toolchains, and named by their vendors, and their composition differs from that of a Linux distribution. RQ4 shows that \TOOLNAME{} recovers many components on proprietary firmware from network security appliances and automotive platforms that existing tools miss, but these images carry no labels, so its recall there cannot be measured. 

\section{Related Work}
\label{sec:related}

Section~\ref{subsec:bg-reference} described how existing detectors identify components in binaries. Here we place \TOOLNAME{} within the wider work on detecting reused components, on generating and assessing SBOMs, and on applying language models to binaries.

\subsection{Detecting Reused Components}
\label{subsec:rw-sca}

Software composition analysis identifies the third-party components of an artifact by matching it against known components, and it has been developed for each kind of artifact. For source code, Centris and OSSFP match the functions of a project against those of large collections of open-source repositories~\cite{woo2021centris,wu2023ossfp}. For Android applications, LibScout matches applications against profiles built from library SDKs~\cite{backes2016libscout}, and ATVHunter distinguishes library versions by control-flow and opcode features drawn from a database of about three million versions~\cite{zhan2021atvhunter}. For native binaries, early systems match string literals and exported function names against package repositories~\cite{hemel2011bat,duan2017osspolice}, and B2SFinder matches binaries against the source code of candidate open-source projects~\cite{yuan2019b2sfinder}. Later systems match learned representations of functions or modules~\cite{tang2022libdb,yang2022modx,li2024libam,jiang2024binaryai,zou2025bincofer} or call graphs~\cite{yu2024dissertation}, adapt the matching to firmware~\cite{zhao2024firmsec}, or identify libraries together with their releases~\cite{dong2024libvdiff,xu2025deeperbin}. A recent study of eleven such tools on 1,873 libraries found an average recall below 60\% and traced the losses to the source features that compilation removes and to the overlap between libraries~\cite{liu2026binarytpl}. These approaches differ in the features they compare, but they share a closed world: a component can be reported only if the reference contains it. \TOOLNAME{} treats identification as an open problem. Its model may name any project, its recall is bounded by the evidence that the binary retains rather than by a reference (Section~\ref{subsec:formulation}), and we evaluate it on components that lie outside every reference in use.

\subsection{SBOM Generation and Quality}
\label{subsec:rw-sbom}

Research on SBOMs has studied how practitioners produce and use them~\cite{empirical,SBOM_lr,bomsaway}, how widely open-source projects adopt them~\cite{nocera2023sbomadoption}, how the choice of generator affects vulnerability reports~\cite{impact}, and how complete and consistent the generated inventories are. Differential analysis of metadata-based generators has revealed inconsistent results and missing dependencies~\cite{yu2024sbomcorrectness}, published Java SBOMs omit dependencies, including vulnerable ones~\cite{xiao2025jbomaudit}, and CovSBOM uses code coverage to tell which declared dependencies are actually used~\cite{covSBOM}. These works start from build or package metadata. For binaries, SBOM tools such as CVE Binary Tool, EMBA, FACT, and Syft identify components with product-specific rules~\cite{cvebintool,emba,fact,syft}, and UniBOM combines binary, file-system, and source analysis to generate SBOMs for IoT firmware and tracks the CPE identifiers of their components over time~\cite{safronov2025unibom}. The identities that these tools report are those of their rules and of vulnerability databases. \TOOLNAME{} adds records for components that no rule and no catalogue entry anticipates, each with its relationship to the file and its evidence, as the SBOM standards provide for~\cite{spdx23,ecma424cyclonedx16}.

\subsection{Language Models for Binary Analysis}
\label{subsec:rw-llm}

Language models are increasingly applied to binaries: they decompile binary code~\cite{tan2024llm4decompile}, analyze code including obfuscated code~\cite{fang2024llmcode}, retrieve the source code from which a binary function was compiled~\cite{gagnon2026anchor}, and act as agents that analyze firmware~\cite{zhang2025firmhive}. For component identification, Pereira et al.\ relate product names to the version strings extracted from binaries by semantic similarity, without training on the target products~\cite{pereira2024sbom}; PyXSieve uses a model to write per-library version-extraction rules that are accepted only if they reproduce known versions~\cite{alexopoulos2026pyxsieve}; BLADE uses a model to select among the candidate libraries that its feature matching retrieves~\cite{liu2026blade}; and our earlier work used a learned matcher between binary strings and product names to identify firmware assets~\cite{ers0}. In these systems, the model works within a reference: a list of candidate products, the libraries whose versions are known, or the candidates that feature matching retrieves. Work on language generation has shown more broadly that model output becomes more reliable when it is revised against retrieved evidence~\cite{gao2023rarr} or verified in a separate step~\cite{dhuliawala2024cove}, and that self-correction without external feedback is unreliable~\cite{huang2024selfcorrect}. \TOOLNAME{} applies this insight to binaries: the model proposes components without a reference, and a deterministic check against the binary's own structure and strings decides which are reported.

\section{Conclusion}
\label{sec:conclusion}

This paper presented \TOOLNAME{}, which recovers SBOM records from ELF binaries without requiring their components to appear in any reference. A language model proposes components and releases from evidence retrieved out of the binary's strings, in a catalogue-guided and an open pass, and a deterministic acceptance step reports a record only when the binary itself supports it, together with the record's relationship to the file and its evidence.

On 1,173 Debian files whose components appear in no reference of the evaluated tools, \TOOLNAME{} recovers the component of 39.6\% of the files, where the existing tools recover at most 0.3\%. It reports the correct release for 44.5\% of the components it recovers, and for 89.3\% when a single string carries both the name and the release, and 78.7\% of its records are correct. The ablation shows that the open pass provides most of this recall and that evidence-gated acceptance keeps the output to about one record per file without discarding the components that it does not accept as embedded. \TOOLNAME{} costs less than half a cent per file and needs no reference to be built in advance. On 101 proprietary firmware images, it reports 148 products that neither CVE Binary Tool nor EMBA reports.

These results show that the evidence left in deployed binaries supports component recovery beyond the references on which existing tools depend. In future work, we plan to reconcile aliases and to model component hierarchies more systematically, to extend the structural evidence beyond ELF, and to integrate binary-derived SBOM recovery into firmware security workflows.

\bibliographystyle{ACM-Reference-Format}
\bibliography{ref.bib}

\end{document}